\documentclass[12pt,prb,aps]{revtex4-1}
\usepackage{amsmath}           		          	
\usepackage{graphicx,epstopdf}					
\usepackage{amssymb}
\usepackage{fullpage}
\usepackage{color}
\usepackage{esint}
\pdfoutput = 1 
\newcommand {\bxi}{\mbox{\boldmath$\xi$}}
\allowdisplaybreaks

\begin{document}
\title{Investigation of Neoclassical Tearing Mode Detection by ECE Radiometry in  an ITER-like Tokamak via Asymptotic Matching Techniques}
\author{Richard Fitzpatrick\,\footnote{rfitzp@utexas.edu}}
\affiliation{Institute for Fusion Studies, Department of Physics, University of Texas at Austin, Austin TX 78712}

\begin{abstract}
The TJ toroidal tearing mode code is used to make realistic predictions of  the electron cyclotron emission (ECE) signals generated by  neoclassical
tearing modes (NTMs) in an ITER-like tokamak plasma equilibrium. In the so-called ``outer region'', which comprises the bulk of the plasma, helical
harmonics of the magnetic field with the same toroidal mode number as the NTM, but different poloidal mode numbers, are coupled together by the Shafranov shift and
shaping of the equilibrium magnetic flux-surfaces. In the ``inner region'', which is localized in the vicinity of the NTM rational surface, helical harmonics whose
poloidal and toroidal mode numbers are in the same ratio as those of the NTM are coupled together nonlinearly to produce a radially asymmetric magnetic
island chain. The solutions in the inner and outer regions are asymptotically matched to one another. The asymptotic matching process determines the
overall magnetic structure of the NTM, as well as the global  perturbation to the electron temperature caused by the mode. A simulated ECE diagnostic is developed that accounts for the
 downshifting and broadening in frequency of the signal due to the relativistic mass increase of the emitting electrons. 

\end{abstract}
\maketitle

\section{Introduction}
The transient heat fluxes and electromagnetic stresses that the plasma facing components would experience during a disruption in
a next-generation tokamak, such as ITER,   that is large enough to produce substantial amounts of fusion energy, are unacceptably large.\cite{iter,wesson}  Consequently, such a tokamak must be capable of  operating reliably in an essentially disruption-free manner. 
Virtually all disruptions in tokamaks are triggered by macroscopic magnetohydrodynamical (MHD) instabilities.\cite{jet,vries} Fortunately, disruptions associated with both ideal instabilities   and
``classical'' tearing instabilities can   be readily avoided  by keeping the toroidal plasma current, the  mean plasma pressure, and
the mean electron number density below  critical values that are either easily calculable or well-known empirically.\cite{iter,decaf}  

A tearing mode\,\cite{tear1} of finite amplitude generates a helical magnetic island chain\,\cite{ntm1} in the vicinity of the ``rational''  magnetic flux-surface\,\cite{ideal3} at which it reconnects magnetic flux.
If the radial width of the island chain exceeds a relatively small threshold value then rapid  heat transport parallel to magnetic field-lines causes a flattening of the  electron temperature profile
that is localized within the chain's magnetic separatrix.\cite{ntm2} The associated loss of the pressure-gradient-driven non-inductive neoclassical bootstrap current\,\cite{ntm3} within the separatrix
has a destabilizing effect that can render a linearly-stable tearing mode unstable at finite amplitude. This type of instability is known as a {\em neoclassical tearing mode}\/  (NTM).\cite{tftr,ntm4c,ntm4b,ntm4a}  A large proportion of fusion-relevant tokamak
discharges are potentially unstable  to 2, 1 and 3, 2 NTMs.\cite{ntm4,ntm5}  (Here, $m, n$ denotes a mode whose resonant harmonic has $m$ periods in the poloidal
direction, and $n$ periods in the toroidal direction.) It is, therefore, not  surprising that NTMs are, by far, the most common cause of disruptions in ITER-baseline-scenario tokamak
discharges.\cite{iter,ntm4,ntm5,vries}
Now, an NTM needs to exceed a critical  threshold amplitude
before it is triggered. In practice, NTMs are
triggered by transient magnetic perturbations associated with other more benign instabilities in the plasma, such as sawtooth oscillations, fishbones, and
edge localized modes (ELMs).\cite{ntm4,ntm5,sawtooth,elm} NTMs pose a unique challenge to next-generation tokamaks  because  virtually all fusion-relevant plasma discharges 
are potentially unstable to multiple NTMs. Moreover, NTM onset is essentially unpredictable, because it
is impossible to determine ahead of time which particular sawtooth crash, fishbone, or ELM is going to trigger a particular NTM.\cite{nstx} Indeed, not all previously documented NTMs possess identifiable
triggers.\cite{ntm6} 

Fortunately, NTMs can be reliably suppressed via electron cyclotron current drive (ECCD).\cite{zohm,prater} This technique, which has been
successfully implemented on many tokamaks,\cite{eccd1,eccd2,eccd3,eccd3a,eccd3b,eccd4,eccd5} involves
launching electron cyclotron waves into the plasma in such a manner that they drive a toroidal current (in the same direction as the equilibrium current) that is 
localized  inside the magnetic separatrix of
the NTM island chain. The idea is to compensate for the loss of the bootstrap current within the separatrix consequent on the local flattening
of the electron temperature profile.\cite{ntm4,ntm5} 

The successful suppression of an NTM via ECCD depends crucially on the {\em early detection}\/ of the mode, combined with an {\em accurate}\/ measurement   of 
the instantaneous location of, at least, one of the O-points  of the associated island chain.\cite{eccd6} In fact, because the island chain is radially thin, 
but relatively extended in poloidal and toroidal angle, the measurement  of the radial location of the O-point is, by far,  the most difficult aspect of
this process. The most convenient method of  detecting an NTM, and simultaneously determining the radial location of an associated island O-point, is to measure the temperature perturbation associated with the mode by means of electron cyclotron emission (ECE) radiometry.\cite{ece1,ece2,ntm2,ece4}

Given the crucial importance of early and accurate detection of NTMs via ECE radiometry to the success of next-generation tokamaks, existing theoretical calculations of
the expected ECE signal are surprisingly primitive.\cite{eccd6,ece4,ece4a} The aim of this paper is to
improve such calculations by taking into account the fact that an NTM  in a realistic toroidal tokamak equilibrium consists of multiple coupled poloidal and toroidal harmonics. Harmonics with the same toroidal mode number as the NTM, but different poloidal mode numbers, are linearly coupled by the
Shafranov shift, elongation, and triangularity of the equilibrium magnetic flux-surfaces.\cite{tear2,tear3,tear5} Furthermore, harmonics whose poloidal and
toroidal mode numbers are in the same ratio as those of the NTM are coupled nonlinearly in the immediate vicinity of the island chain.\cite{ntm1,ntm2}
Previous calculations have taken into account the important fact that an NTM island chain is likely to be radially asymmetric with respect to the rational surface,\cite{ece6a,ece6} 
due to the mean radial plasma displacement at the  surface, but have not necessarily made an accurate
determination of this asymmetry.\cite{eccd6} Our improved calculation incorporates an accurate assessment of the asymmetry. Finally, the ECE signal 
is downshifted and broadened in frequency due to the relativistic mass increase of the emitting electrons.\cite{ece1,ece2,ece5}  This process leads to a shift in the inferred location
of the ECE  to larger major radius, as well as a radial smearing out the emission. Both of these effects, which limit the accuracy to which the
radial location of the island O-point can be measured via ECE,  are taken into account in our improved calculation.

The calculation of the  magnetic perturbation associated with an NTM is most efficiently formulated as an {\em asymptotic matching}\/ problem in which the  plasma is  divided into two distinct regions.\cite{tear1,tear2,tear3,tear4,tear5,tear6,tear7,tear8,tear9,tear10}    In the so-called {\em outer region}, which comprises most
of the plasma, the perturbation is governed by the equations of linearized, marginally-stable, ideal-MHD.
However, these equations become singular on   {\em rational}\/ magnetic flux-surfaces at which the perturbed magnetic field resonates with the equilibrium field. In the {\em inner region}, which
consists of a set of narrow layers centered on the various rational surfaces, non-ideal-MHD effects such as plasma resistivity, as well as nonlinear effects,  become important. 
 In the calculation described in this paper, the NTM is assumed to reconnect magnetic flux at one particular rational surface in the plasma (i.e., the
 $q=2$ surface for the case of a 2, 1 mode, and the $q=3/2$ surface in the case of a 3, 2 mode). The response of the plasma at the
 other rational surfaces is assumed to be ideal, as we would expect to be the case in the presence of sheared plasma rotation.\cite{tear5}
The magnetic perturbation in the segment of the inner region centered on the reconnecting rational surface is that associated with a radially asymmetric magnetic island chain.\cite{ntm1,island}
The nonlinear island solution needs to be asymptotically matched to the linear ideal-MHD solution in the outer region. The
electron temperature perturbation associated with the NTM in the inner and outer regions is simultaneously  determined by the asymptotic matching process. 

In this paper, the asymptotic matching is performed using the TJ toroidal tearing mode code,\cite{tear9,tear10}  which employs an aspect-ratio
expanded toroidal magnetic equilibrium.\cite{exp} The TJ code is used for the sake of convenience. However, the calculations described in this paper
could just as well be implemented using a toroidal tearing mode code, such as STRIDE,\cite{tear7,tear8} that employs a general toroidal magnetic 
equilibrium. 

This paper is organized as follows. The adopted plasma equilibrium is described in Sect.~\ref{s2}.  In Sect.~\ref{s3}, the perturbed electron temperature associated with  an NTM is calculated in the outer region. The corresponding perturbed electron temperature  in the inner region is calculated in Sect.~\ref{s4}. The
global perturbed electron temperature, obtained by asymptotically matching the perturbed temperatures  in  the inner and outer regions, is
described in Sect.~\ref{s5}. 
The ECE signal due to the NTM is calculated in Sect.~\ref{s6}.
 Finally, the paper is summarized, and conclusions are drawn, in Sect.~\ref{s7}. 

\section{Model Plasma Equilibrium}\label{s2}
\subsection{Normalization}\label{norm}
Unless otherwise specified, all lengths in this paper are normalized to  the major radius of the plasma magnetic axis, $R_0$, All  magnetic field-strengths
are normalized to the  toroidal field-strength at the magnetic axis, $B_0$. All plasma pressures are normalized to $B_0^{\,2}/\mu_0$.

\subsection{Coordinates}\label{coord}
Let $R$, $\phi$, $Z$ be right-handed cylindrical coordinates whose symmetry axis corresponds to the symmetry axis of the axisymmetric toroidal plasma equilibrium.
Let $r$, $\theta$, $\phi$ be right-handed flux-coordinates whose
Jacobian is
\begin{equation}\label{jac}
{\cal J}(r,\theta)\equiv (\nabla r\times \nabla\theta\cdot\nabla\phi)^{-1}= r\,R^{\,2}.
\end{equation}
Note that $r=r(R,Z)$ and $\theta=\theta(R,Z)$. 
The magnetic axis corresponds to $r=0$, and the plasma-vacuum interface to $r=a$. Here, $a\ll 1$ is the effective inverse aspect-ratio of the plasma. 

\subsection{Equilibrium Magnetic Field}\label{equilb}
Consider a tokamak plasma equilibrium whose magnetic field takes the form
\begin{equation}
{\bf B}(r,\theta) = f(r)\,\nabla\phi\times \nabla r + g(r)\,\nabla\phi = f\,\nabla(\phi-q\,\theta)\times \nabla r,
\end{equation}
where
$q(r) = r\,g/f$ is the safety-factor profile.
Equilibrium force balance requires that
$ \nabla P={\bf J}\times {\bf B}$, 
where 
\begin{equation}
P(r)= a^2\,p_2(r),
\end{equation}
 is the equilibrium scalar plasma pressure profile, and ${\bf J}=\nabla\times {\bf B}$ the equilibrium plasma current density. 
 The (unnormalized) equilibrium electron temperature profile is written
 \begin{equation}
 T_{e\,0}(r) = \frac{B_0^{\,2}}{\mu_0}\,\frac{P(r)}{2\,n_e(r)} + T_{e\,{\rm ped}},
 \end{equation}
 where $n_e(r)$ is the (unnormalized) equilibrium electron number density profile. Here, we are assuming that the electrons and ions have the same
 temperature, as is likely to be the case in ITER. 

\subsection{Equilibrium Magnetic Flux-Surfaces}
The loci of the up-down symmetric equilibrium magnetic flux-surfaces are written in the parametric form\,\cite{tear5}
\begin{align}
R(\hat{r},\omega) &= 1 -a\,\hat{r}\,\cos\omega + a^{2}\left[H_1(\hat{r})\,\cos \omega + H_2(\hat{r})\,\cos 2\,\omega+H_3(\hat{r})\,\cos 3\,\omega\right], \label{e19x}\\[0.5ex]
Z(\hat{r},\omega)&= a\,\hat{r}\,\sin\omega +a^{2}\left[H_2(\hat{r})\,\sin 2\,\omega+H_3(\hat{r})\,\sin 3\,\omega\right], \label{e20x}
\end{align}
where  $r=a\,\hat{r}$. 
Here, the dimensionless functions $H_1(\hat{r})$, $H_2(\hat{r})$, and $H_3(\hat{r})$ control the Shafranov shift, vertical elongation, and  triangularity of
the flux-surfaces, respectively. 
Moreover,\cite{exp}
\begin{align}
g(\hat{r}) &= 1+ a^2\,g_2(\hat{r}),\\[0.5ex]
g_2'&= -p_2' - \frac{\hat{r}}{q^2}\,(2-s),\\[0.5ex]
H_1''&= -(3-2\,s)\,\frac{H_1' }{\hat{r}}-1+\frac{2\,p_2'\,q^2}{\hat{r}},\label{e27}\\[0.5ex]
H_j''&= -(3-2\,s)\,\frac{H_j'}{\hat{r}}+(j^2-1)\,\frac{H_j}{\hat{r}^{\,2}}~~~~~\mbox{for $j>1$},\label{e33x}\\[0.5ex]
\theta &= \omega+a\,\hat{r}\,\sin\omega - a\sum_{j=1,3}\frac{1}{j}\left[H_j'-(j-1)\,\frac{H_j}{\hat{r}}\right]\sin j\,\omega,
\end{align}
where $s(\hat{r}) = \hat{r}\,q'/q$ is the magnetic shear, and $'$ denotes $d/d\hat{r}$. The plasma equilibrium is fully specified by the value of $a$, the two free
flux-surface functions $q(\hat{r})$ and $p_2(\hat{r})$, the values of $H_2(1)$ and $H_3(1)$, and the electron number density profile, $n_e(\hat{r})$. 

\subsection{Example Plasma Equilibrium}
\begin{figure}
\centerline{\includegraphics[width=0.85\textwidth]{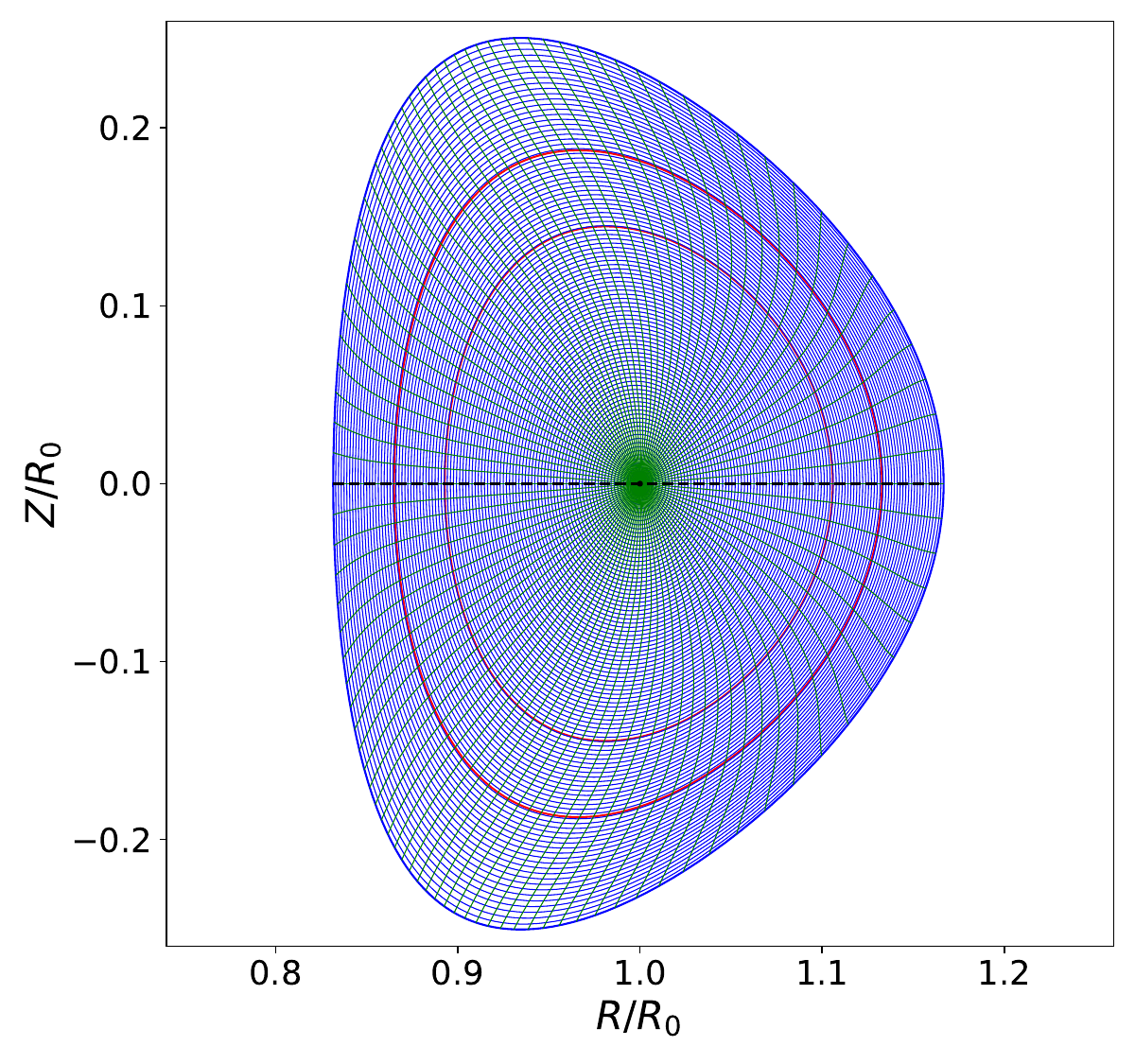}}
\caption{The blue and green curves show surfaces of constant $r$ and $\theta$, respectively, for an ITER-like example
plasma equilibrium characterized by $B_0=5.3$ T, $R_0=6.2$ m, $a=0.2$, $H_2(1)=1.0$, and $H_3(1)= 0.5$. The red curves show
the locations of the $q=3/2$ and $q=2$ surfaces. The black dot shows the location of the magnetic axis. The dashed black line shows the location of the ECE measurement chord. \label{figa}}
\end{figure}

\begin{figure}
\centerline{\includegraphics[width=1.0\textwidth]{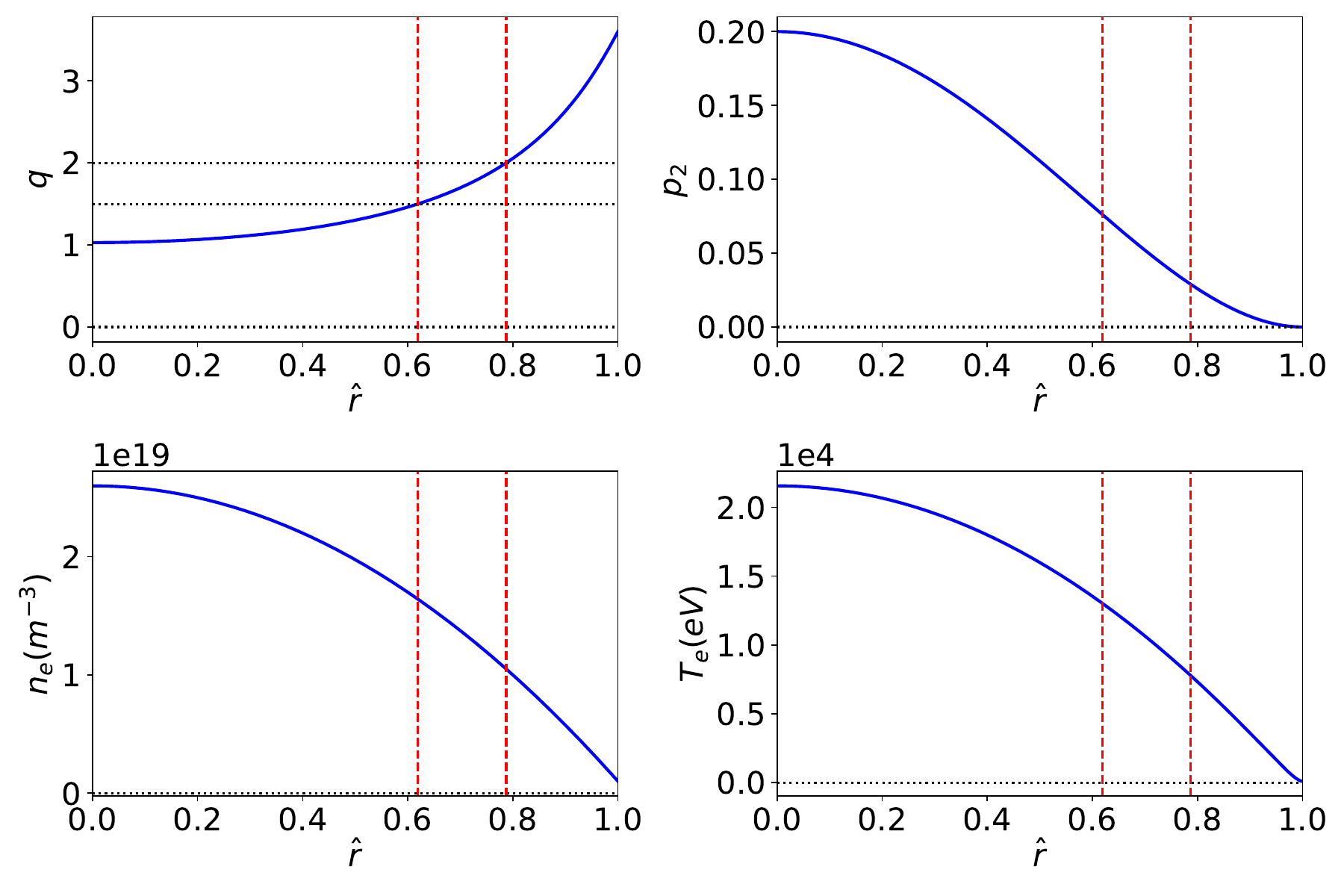}}
\caption{The safety-factor, normalized pressure, electron number density, and electron temperature profiles for the
example equilibrium pictured in Fig.~\ref{figa}. The vertical red lines show the locations of the $q=3/2$ and $q=2$
surfaces.   \label{figb}}
\end{figure}

Figures~\ref{figa} and \ref{figb} show the magnetic flux-surfaces and profiles of an ITER-like  example plasma equilibrium characterized by 
$B_0=5.3$ T, $R_0=6.2$ m, $a=0.2$, $H_2(1)=1.0$,  $H_3(1)= 0.5$, $q(0)=1.01$, and  $q(a)=3.6$.  The normalized plasma inductance is
$l_i=1.16$, and the normalized beta is $\beta_N= 1.32$.
The toroidal plasma current is $I_t=2.44$\,MA. This rather low value
is symptomatic of the fact that the true ITER inverse aspect-ratio is  $0.32$ rather than $0.20$. Unfortunately, the TJ code, which is based on
an expansion in the inverse aspect-ratio,  does not give accurate results for inverse aspect-ratios
in excess of 0.2. 
Note that the normalized central pressure, $p_2(0)$---and, hence, the product of the central electron number density and the electron temperature---is  limited by the requirement that the classical
 2,\,1 and 3, 2 tearing modes remain stable. 
Thus,  a rather small value of the central electron number density, $n_{e\,0}=2.5\times 10^{19}\,{\rm m}^{-3}$, 
must be adopted  in order to permit a realistic central electron number density of $T_{e\,0}\sim 20\,{\rm keV}$. 

\section{Perturbed Electron Temperature  in Outer Region}\label{s3}
\subsection{Perturbation in Outer Region}
Let the positive integer $n$ be the toroidal mode number of the NTM. Let there be $K$ rational surfaces in the plasma, of minor radius $r_k$ (for $k=1,K$),  at which the resonance condition
$q(r_k) = m_k/n$ is satisfied, where the positive integer $m_k$ is the resonant poloidal mode number at the $k$th surface. The perturbed magnetic field in the outer region is specified by\,\cite{tear9,tear10}
\begin{equation}
b^r(r,\theta,\phi)\equiv {\bf b}\cdot\nabla r = \frac{{\rm i}}{r\,R^{\,2}}\sum_{j=1,J} \psi_{m_j}(r)\,{\rm e}^{\,{\rm i}\,(m_j\,\theta-n\,\phi)}.
\end{equation}
Here, ${\bf b}(r,\theta,\phi)$ is the perturbed magnetic field-strength, and the $m_j$ are the  $J>K$ poloidal mode numbers included in the calculation. (The $m_k$ are a subset of the $m_j$.) 

The functions $\psi_{m_j}(r)$ are determined by solving a set of $2\,J$ coupled ordinary differential equations that are singular at the
various rational surfaces in the plasma. The solutions to these equations must be launched from the magnetic axis ($r=0$), integrated outward in $r$, stopped   just before and  restarted just after each rational surface in the plasma, integrated to the plasma boundary ($r=a$), and then matched to a free-boundary vacuum
solution. This process is described in detail in Ref.~\onlinecite{tear9}. 

\subsection{Behavior in Vicinity of Rational Surface}\label{rational}
Consider the behavior of the $\psi_{m_j}(r)$ in the vicinity of the $k$th rational surface. 
The non-resonant $\psi_{m_j}(r$), for which $m_j\neq m_k$,   are continuous across the surface. On the other hand, the resonant $\psi_{m_j}(r)$ is
such that
\begin{align}\label{rat}
\psi_{m_k}(r_k+x) &= A_{L\,k}\,|x|^{\nu_{L\,k}} + {\rm sgn}(x)\,A_{S\,k}^{\pm}\,|x|^{\nu_{S\,k}},\\[0.5ex]
\end{align}
where
\begin{align}
\nu_{L\,k} &= \frac{1}{2}-\sqrt{-D_{I\,k}},\\[0.5ex]
\nu_{S\,k} &= \frac{1}{2}+\sqrt{-D_{I\,k}},\\[0.5ex]
D_{I\,k}&= - \left[\frac{2\,(1-q^2)}{s^2}\,r\,\frac{dP}{dr}\right]_{r_k} -\frac{1}{4}\label{di}
\end{align}
Here,  $A_{L\,k}$ is termed the coefficient of the {\em large}\, solution, whereas $A_{S\,k}^\pm$ are the coefficients of the {\em small}\/ solution.
Here, $A_{S\,k}^+$ corresponds to the region $x>0$, whereas  $A_{S\,k}^-$ corresponds to the region $x<0$. Furthermore, $D_{I\,k}$ is the ideal
Mercier interchange parameter (which needs to be negative to ensure stability to localized interchange modes),\cite{mercier,ggj,ggj1} and $\nu_{L\,k}$ and $\nu_{S\,k}$
are termed the {\em Mercier indices}. 

It is helpful to define the quantities\,\cite{tear9}
\begin{align}\label{Psidef}
{\mit\Psi}_k&= r_k^{\,\nu_{L\,k}}\left(\frac{\nu_{S\,k}-\nu_{L\,k}}{L_{m_k}^{\,{m_k}}}\right)^{1/2}_{r_k} A_{L\,k},\\[0.5ex]
{\mit\Delta\Psi}_k &= r_k^{\,\nu_{S\,k}}\left(\frac{\nu_{S\,k}-\nu_{L\,k}}{L_{m_k}^{\,m_k}}\right)^{1/2}_{r_k} (A_{S\,k}^+-A_{S\,k}^-),\label{edpp}
\end{align}
at each rational surface in the plasma, where
\begin{align}
L_{m_k}^{\,m_k}(r) &= m_k^2\,c_{m_k}^{\,m_k}(r) + n^2\,r^2,\\[0.5ex]
c_{m_k}^{\,m_k}(r) &=\oint|\nabla r|^{-2}\,\frac{d\theta}{2\pi}.
\end{align}
 Here, the dimensionless complex parameter ${\mit\Psi}_k$ is a measure of the reconnected helical magnetic flux at the $k$th rational surface, whereas
the dimensionless complex parameter ${\mit\Delta\Psi}_k$ is a measure of the strength of a localized current sheet that flows parallel to the equilibrium magnetic field at the surface.

 It is assumed that ${\mit\Psi}_k = 0$ for all $k$, except for $k=l$. In other words, the NTM only reconnects magnetic flux at the
$l$th rational surface. Let 
\begin{equation}
\psi_{m_j}(r) = {\mit\Psi}\,\hat{\psi}_{m_j}(r),
\end{equation}
where ${\mit\Psi}$ is the reconnected magnetic flux at the $l$th rational surface, and the $\hat{\psi}_{m_j}(r)$ are normalized such that ${\mit\Psi}_k
= \delta_{kl}$. 

\subsection{Electron Temperature in Outer Region}
Let $\bxi(r,\theta,\phi)$ be the plasma displacement in the outer region. We can write\,\cite{tear10}
\begin{align}\label{xi}
\xi^r(r,\theta,\phi)\equiv \bxi\cdot\nabla r& = \sum_{j=1,J}\xi_{m_j}^r(r)\,{\rm e}^{\,{\rm i}\,(m_j\,\theta-n\,\phi)}\nonumber\\[0.5ex]
&=
{\mit\Psi}\,\frac{q}{r\,g}\sum_{j=1,J}\frac{\hat{\psi}_{m_j}}{m_j-n\,q}\, {\rm e}^{\,{\rm i}\,(m_j\,\theta-n\,\phi)}.
\end{align}
The perturbed electron temperature  in the outer region is written
\begin{equation}
\delta T_e(r,\theta,\phi)= -\frac{dT_{e\,0}}{dr}\, \xi^r(r,\theta,\phi) + \delta T_{e\,0}\,H(r-r_{l})
\end{equation}
where
\begin{equation}
H(x)= \left\{\begin{array}{ccc}1&~~~&x<0\\0&&x>0\end{array}\right..
\end{equation}
Here, we are assuming that the electron temperature is passively convected by the plasma in the outer region. We are also
assuming that there is no change in topology of the magnetic flux-surfaces in the outer region. In other words, any topology changes are
confined to  the inner region. Finally, $\delta T_{e\,0}<0$ is the reduction in the equilibrium electron temperature in the plasma
core due to the flattening of the temperature profile in the vicinity of the NTM island chain.\cite{chang}

\section{Perturbed Electron Temperature  in Inner Region}\label{s4}
\subsection{Introduction}
Consider the segment of the inner region in the vicinity of the $l$th rational surface, where the NTM reconnects magnetic flux. 
Let $x=r-r_{l}$, $X=x/W$, and $\zeta=m_{l}\,\theta-n\,\phi$, where $W\ll a$ is the full width  of the NTM island chain's magnetic separatrix. 
Here, $m_l$ is the resonant poloidal mode number at the $l$th rational surface. 
Let us search for a single-helicity solution in which the magnetic flux-surfaces in the vicinity of the island chain are contours of some function ${\mit\Omega}(X,\zeta)$.
Now, a magnetic island chain whose width exceeds the linear layer width is a helical magnetic equilibrium.\cite{ntm1} As such, the island magnetic-flux surfaces must satisfy the fundamental
force balance requirement\,\cite{island}
\begin{equation}\label{e26}
\left[\left.\frac{\partial^2{\mit\Omega}}{\partial X^2}\right|_\zeta,{\mit\Omega}\right]=0,
\end{equation}
where
\begin{equation}\label{poisson}
[A,B] \equiv \left.\frac{\partial A}{\partial X}\right|_\zeta \left.\frac{\partial B}{\partial\zeta}\right|_X- \left.\frac{\partial B}{\partial X}\right|_\zeta \left.\frac{\partial A}{\partial\zeta}\right|_X.
\end{equation}
This requirement stipulates that the current density in the island region must be constant on magnetic flux-surfaces.\cite{ntm1,multi}

\subsection{Island Magnetic Flux-Surfaces}
\begin{figure}
\centerline{\includegraphics[width=0.9\textwidth]{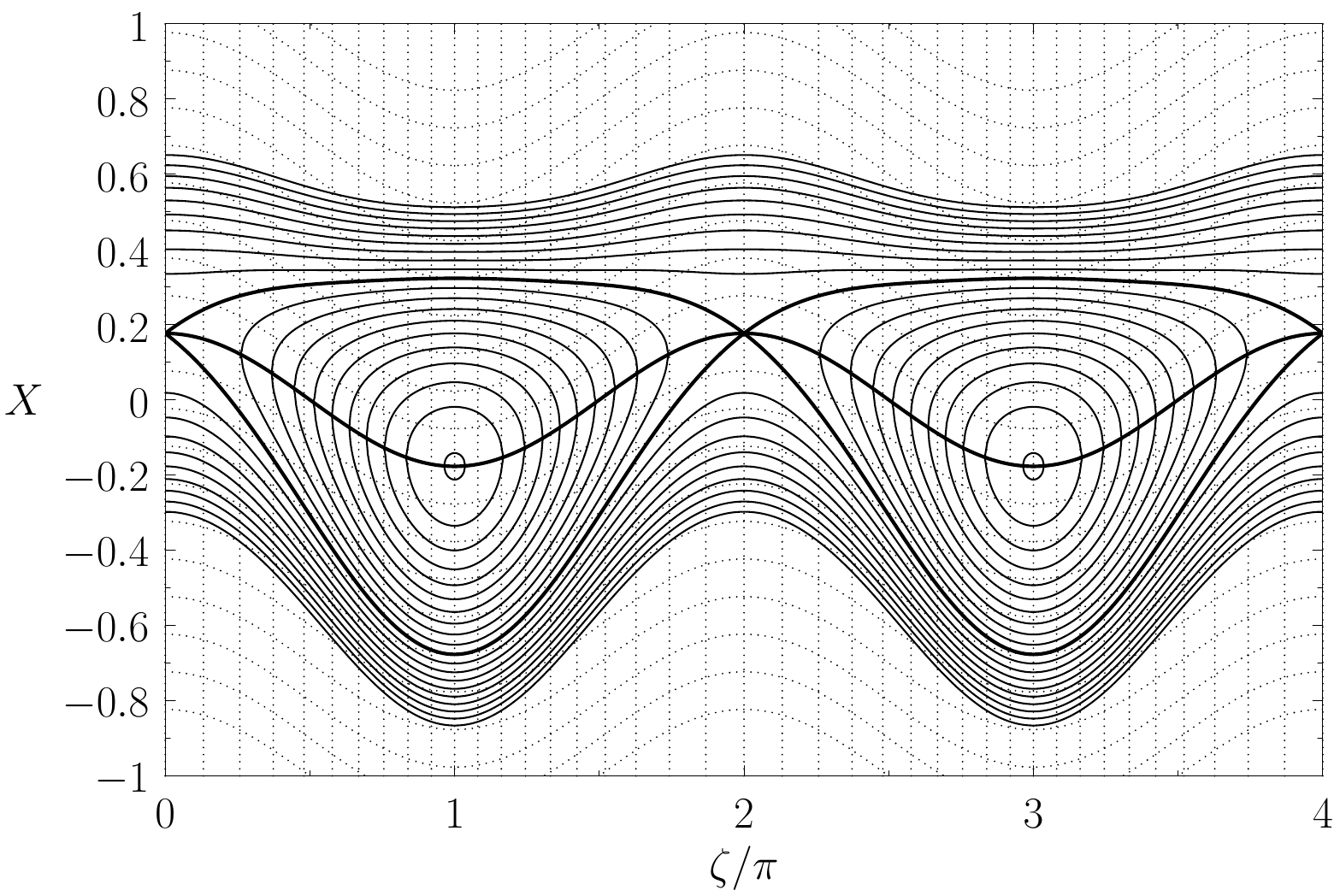}}
\caption{The thin solid curves show the contours of ${\mit\Omega}(X,\zeta)$ evaluated for $\delta=0.5$. The thick solid
lines show the magnetic separatrix (upper and lower curves) and the contour $Y=0$ (middle curve). The curved dotted
lines show equally-spaced contours of $Y$, whereas the vertical dotted lines show equally-spaced contours of $\xi$. 
(Reproduced, with permission, from Ref.~\onlinecite{island}.) \label{fig1}}
\end{figure}

A suitable solution of Eq.~(\ref{e26}) that connects to the ideal-MHD solution in the outer region is\,\cite{island}
\begin{equation}\label{e45}
{\mit\Omega}(X,\zeta) = 8\,X^2 + \cos(\zeta-\delta^2\,\sin\zeta) - 2\sqrt{8}\,\delta\,X\,\cos\zeta+\delta^2\,\cos^2\zeta,
\end{equation}
where $|\delta|<1$.  As illustrated in Fig.~\ref{fig1}, the magnetic flux-surfaces  (i.e., the contours of ${\mit\Omega}$) map out an
asymmetric (with respect to $X=0$) island chain whose 
X-points lie at $X=\delta/\sqrt{8}$, $\zeta = 0$, $2\pi$, and ${\mit\Omega}=+1$,  and whose  O-points lie at
$X=-\delta/\sqrt{8}$,  $\zeta=\pi$, and ${\mit\Omega}=-1$. The maximum width of the magnetic separatrix (in $x$) is $W$. 

The first term on the right-hand side of Eq.~(\ref{e45}) emanates from the unperturbed (by the NTM) plasma equilibrium, whereas the
remaining terms emanate from the NTM perturbation in the outer region. In particular, the third term on the right-hand
side, which governs the island asymmetry,  originates from  the mean radial gradient in the $\cos\zeta$ component of the linear NTM eigenfunction at the rational surface. 

The island asymmetry is characterized by the dimensionless parameter $\delta$. If $\delta >0$ then the island O-points are displaced radially inward (with respect to the unperturbed rational
surface), whereas the X-points are displaced radially outward an equal distance. The opposite is the case if $\delta <0$. Generally speaking,  we
expect $\delta>0$ for NTMs (because the linear eigenfunctions for such modes tend to attain their
maximum amplitudes inside the rational surface; see Fig.~2 in Ref.~\onlinecite{white}). Note that if  $|\delta|$ exceeds the critical value
unity then the X-points bifurcate, and a current sheet forms between them.\cite{wal}  Consequently,  it is no longer possible to analyze the resistive evolution of the  resulting island chain  using a variant of standard Rutherford island theory.\cite{ntm1} Hence, we shall only consider the case $-1\leq \delta < 1$. 

\subsection{Coordinate Transformation}
\begin{figure}
\centerline{\includegraphics[width=0.9\textwidth]{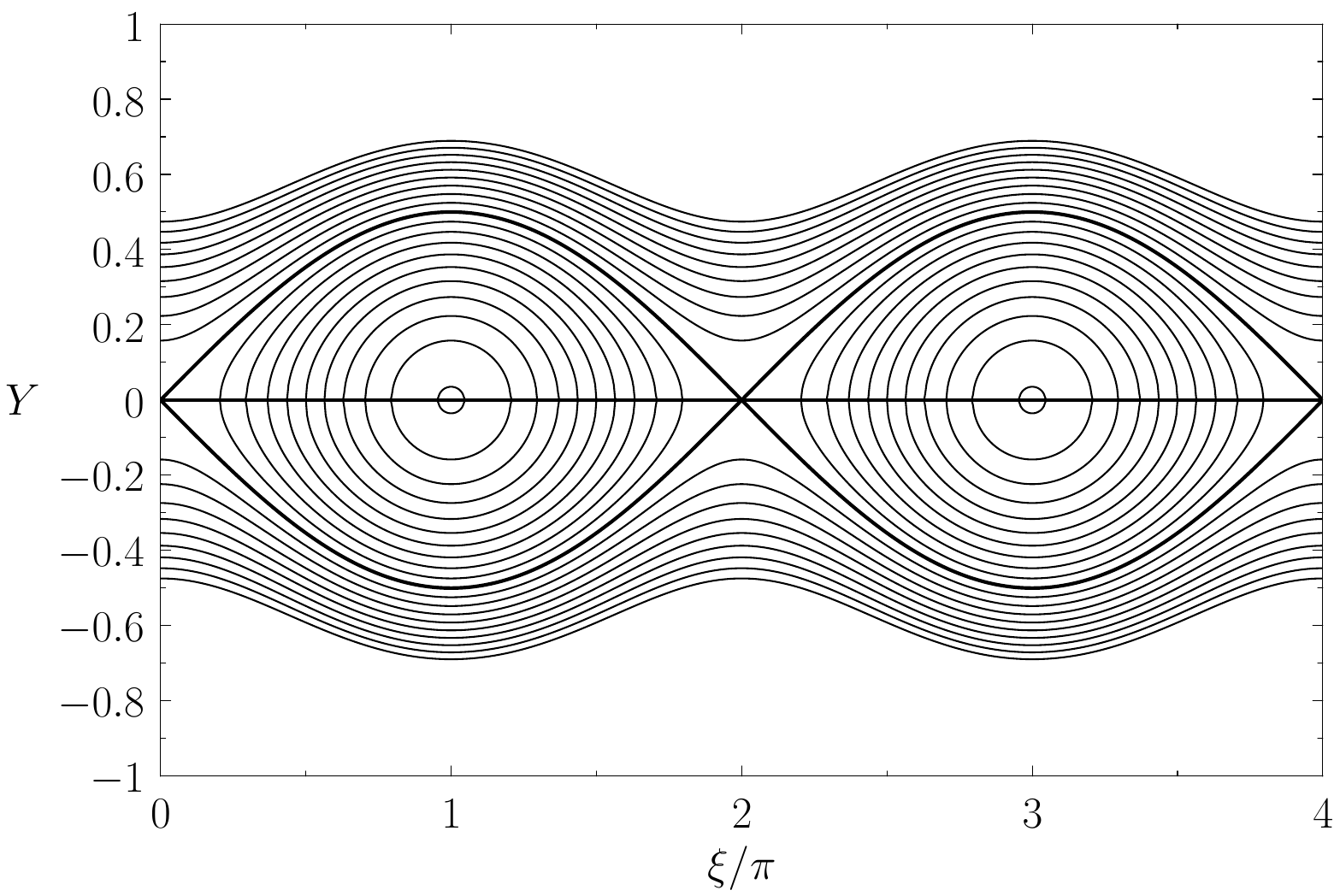}}
\caption{The thin solid curves show the contours of ${\mit\Omega}(Y,\xi)$ evaluated for $\delta=0.5$. The thick solid
lines show the magnetic separatrix (upper and lower curves) and the contour $Y=0$ (middle curve). (Reproduced, with permission, from Ref.~\onlinecite{island}.) \  \label{fig2}}
\end{figure}

Let us define the new coordinates\,\cite{island}
\begin{align}
Y &= X-\frac{\delta}{\sqrt{8}}\,\cos\zeta,\label{ek}\\[0.5ex]
\xi&=\zeta-\delta^2\,\sin\zeta.\label{ekepler}
\end{align}
When expressed in terms of these coordinates, the magnetic flux-function (\ref{e45}) reduces to the simple
form
\begin{equation}\label{eeven}
{\mit\Omega}(Y,\xi) =8\,Y^{\,2}+\cos\xi.
\end{equation}
Thus, as illustrated in Fig.~\ref{fig2}, irrespective of the value of the asymmetry parameter, $\delta$, when
plotted in $Y$, $\xi$ space, the magnetic flux-surfaces map out a symmetric (with respect to $Y=0$) island
chain whose O-points lie at $\xi=\pi$, $Y=0$,
and ${\mit\Omega}=-1$, and whose X-points lie at $\xi=0$, $2\pi$, $Y=0$, and ${\mit\Omega}=+1$.  The fact that the radially asymmetric island flux-surfaces can
be remapped to a set of radially symmetric surfaces greatly simplifies our analysis. 

The inversion of Eq.~(\ref{ekepler}) is very well known:\,\cite{bc}
\begin{align}\label{e9a1}
\zeta &= \xi + 2\sum_{\mu=1,\infty}\left[\frac{J_\mu(\mu\,\delta^2)}{\mu}\right]\sin(\mu\,\xi),\\[0.5ex]
\cos\zeta &=-\frac{\delta^2}{2}+\sum_{\mu=1,\infty}\left[\frac{J_{\mu-1}(\mu\,\delta^2)-J_{\mu+1}(\mu\,\delta^2)}{\mu}\right]\cos(\mu\,\xi),\\[0.5ex]
\sin\zeta &=\frac{2}{\delta^2}\sum_{\mu=1,\infty}\left[\frac{J_\mu(\mu\,\delta^2)}{\mu}\right]\sin(\mu\,\xi),\label{e9a3}\\[0.5ex]
\cos(\nu\,\zeta)&= \nu\sum_{\mu=1,\infty}\left[\frac{J_{\mu-\nu}(\mu\,\delta^2)-J_{\mu+\nu}(\mu\,\delta^2)}{\mu}\right]\cos(\mu\,\xi),\\[0.5ex]
\sin(\nu\,\zeta)&= \nu\sum_{\mu=1,\infty}\left[\frac{J_{\mu-\nu}(\mu\,\delta^2)+J_{\mu+\nu}(\mu\,\delta^2)}{\mu}\right]\cos(\mu\,\xi),\
\end{align}
for $\nu>1$. 

\subsection{Plasma Displacement}
Outside the magnetic separatrix, we can write
\begin{equation}
{\mit\Omega}(X,\zeta) = 8\,(X-{\mit\Xi})^2,
\end{equation}
where ${\mit\Xi}= \xi^r/W$ is the normalized radial plasma displacement. It follows that, in the limit $|X|\gg 1$, 
\begin{align}
{\mit\Xi}(X,\zeta)&= -\frac{[{\mit\Omega}(X,\zeta)-8\,X^2 - 8\,{\mit\Xi}^2]}{16\,X}\nonumber\\[0.5ex]
&=\frac{\delta}{\sqrt{8}}\,\cos\zeta - \frac{\cos(\zeta-\delta^2\,\sin\zeta) +\delta^2\,\cos^2\zeta}{16\,X}+ \frac{{\mit\Xi}^2}{2\,X}\nonumber\\[0.5ex]
&\simeq \frac{\delta}{\sqrt{8}}\,\cos\zeta- \frac{\cos(\zeta-\delta^2\,\sin\zeta)}{16\,X},
\end{align}
where use has been made of Eq.~(\ref{e45}).
Note that ${\mit\Xi}(X,\zeta)$ is an even function of $\zeta$. 
Let us write
\begin{equation}
{\mit\Xi}(X,\zeta)= \sum_{\nu=0,\infty} {\mit\Xi}_\nu(X)\,\cos(\nu\,\zeta).
\end{equation}
Thus,
\begin{align}
{\mit\Xi}_1(X) =2 \oint {\mit\Xi}(X,\zeta)\,\cos(\zeta)\,\frac{d\zeta}{2\pi} &= \frac{\delta}{\sqrt{8}}-\frac{1}{8\,X}\oint\cos(\zeta-\delta^2\,\sin\zeta)\,\cos\zeta\,\frac{d\zeta}{2\pi}\nonumber \\[0.5ex]
&=\frac{\delta}{\sqrt{8}} -\frac{1}{16\,X}\oint\cos(-\delta^2\,\sin\zeta)\,\cos\zeta\,\frac{d\zeta}{2\pi}\nonumber\\[0.5ex]
&\phantom{=}- \frac{1}{16\,X}\oint\cos(2\,\zeta-\delta^2\,\sin\zeta)\,\cos\zeta\,\frac{d\zeta}{2\pi}.
\end{align}
But,\cite{bc,grx}
\begin{equation}
J_\nu(\delta^2) = \oint\cos(\nu\,\zeta-\delta^2\,\sin\zeta)\,\frac{d\zeta}{2\pi},
\end{equation}
so
\begin{equation}
{\mit\Xi}_1(X) =\frac{\delta}{\sqrt{8}} - \frac{J_0(\delta^2) + J_2(\delta^2)}{16\,X},
\end{equation}
and
\begin{equation}\label{ea}
\xi^r_1(r_{l}+x) = \frac{W\,\delta}{\sqrt{8}} - \frac{W^2}{16\,x}\,[J_0(\delta^2)+ J_2(\delta^2)].
\end{equation}

In the outer region,  $\xi_{m_{l}}^r(r)$ is the equivalent quantity to $\xi_1^r(r)$.  It follows from Eq.~(\ref{xi}) that, in the limit $|x|\ll a$, 
\begin{equation}\label{eb}
\xi_{m_{l}}^r(r_{l}+x)={\mit\Psi}\,\frac{q}{r\,g}\,\frac{\hat{\psi}_{m_{l}}}{m_{l}-n\,q}
= -{\mit\Psi}\left(\frac{h\,q}{s\,q}\right)_{r_l}\frac{1}{x} + {\cal O}(1),
\end{equation}
where 
\begin{equation}
h(r) = \frac{(L_{m_{l}}^{m_{l}})^{1/2}}{m_{l}},
\end{equation}
and use has been made of Eqs.~(\ref{rat}) and (\ref{Psidef}). Here, we are assuming that $\nu_{L\,l}\simeq 0$ and $\nu_{S\,l}\simeq 1$,
as is generally the case in a large aspect-ratio tokamak. 
A comparison between Eqs.~(\ref{ea}) and (\ref{eb}) reveals that
\begin{equation}\label{e46}
{\mit\Psi} = \left(\frac{W}{4}\right)^2\left(\frac{s\,g}{h\,q}\right)_{r_{l}}\left[J_0(\delta^2)+ J_2(\delta^2)\right],
\end{equation}
and
\begin{equation}\label{e47}
\delta \simeq \frac{\sqrt{2}}{W}\left[\xi_{m_{l}}^r(r_{l}+W)+\xi_{m_{l}}^r(r_{l}-W)\right].
\end{equation}

Equation~(\ref{e46}) gives the relationship between the reconnected magnetic flux, ${\mit\Psi}$, and the island width, $W$. This relationship
differs from the conventional one\,\cite{ntm1} because of corrections due to the radial asymmetry of the island chain. However, the corrections are fairly
minor. In fact, $1\geq J_0(\delta^2)+ J_2(\delta^2)\geq 0.880$ for $0\leq |\delta|\leq 1$. Equation~(\ref{e47}) specifies the relationship between the
island asymmetry parameter, $\delta$,  and the mean radial plasma displacement at the rational surface. Note that the matching
between the inner and outer solutions is performed at $r=r_{l}\pm W$. 

\subsection{Flux-Surface Average Operator}
Now, 
\begin{equation}\label{e49}
\left.\frac{\partial}{\partial X}\right|_\zeta= \left.\frac{\partial{\mit\Omega}}{\partial X}\right|_\zeta\left.\frac{\partial}{\partial{\mit\Omega}}\right|_{\xi}+ \left.\frac{\partial\xi}{\partial X}\right|_\zeta\left.\frac{\partial}{\partial\xi}\right|_{\mit\Omega}=16\,Y\left.\frac{\partial}{\partial{\mit\Omega}}\right|_{\xi},
\end{equation}
and
\begin{equation}
\left.\frac{\partial}{\partial \zeta}\right|_X= \left.\frac{\partial{\mit\Omega}}{\partial \zeta}\right|_X\left.\frac{\partial}{\partial{\mit\Omega}}\right|_{\xi}+ \left.\frac{\partial\xi}{\partial \zeta}\right|_X\left.\frac{\partial}{\partial\xi}\right|_{\mit\Omega},
\end{equation}
so
\begin{equation}
[A,B] \equiv \frac{16\,Y}{\sigma}\left(\left.\frac{\partial A}{\partial{\mit\Omega}}\right|_\xi\left.\frac{\partial B}{\partial\xi}\right|_{\mit\Omega}-\left.\frac{\partial B}{\partial{\mit\Omega}}\right|_\xi\left.\frac{\partial A}{\partial\xi}\right|_{\mit\Omega}\right),
\end{equation}
where
\begin{equation}\label{sigma}
\sigma(\xi) \equiv\frac{d\zeta}{d\xi}=  1+2\sum_{\mu=1,\infty} J_\mu(\mu\,\delta^2)\,\cos(\mu\,\xi),
\end{equation}
and use has been made of Eqs.~(\ref{poisson})--(\ref{ekepler}) and (\ref{e9a1}). 
In particular,
\begin{equation}\label{e52}
[A,{\mit\Omega}] = -\frac{16\,Y}{\sigma}\left.\frac{\partial A}{\partial\xi}\right|_{\mit\Omega}.
\end{equation}

The flux-surface average operator, $\langle\cdots\rangle$, is the annihilator of $[A,{\mit\Omega}]$ for arbitrary $A(\varsigma,{\mit\Omega},\xi)$.\cite{ntm2,island} Here, $\varsigma=+1$ for $Y>0$ and $\varsigma =-1$ for
$Y<0$. It follows from Eq.~(\ref{e52}) that
\begin{equation}
\langle A\rangle = \int_{\zeta_0}^{2\pi-\zeta_0}\frac{\sigma(\xi)\,A_+({\mit\Omega},\xi)}{\sqrt{2\,({\mit\Omega}-\cos\xi)}}\,\frac{d\xi}{2\pi}
\end{equation}
for $-1\leq {\mit\Omega}\leq 1$, and
\begin{equation}\label{e55}
\langle A\rangle = \int_0^{2\pi}\frac{\sigma(\xi)\,A(\varsigma,{\mit\Omega},\xi)}{\sqrt{2\,({\mit\Omega}-\cos\xi)}}\,\frac{d\xi}{2\pi}
\end{equation}
for ${\mit\Omega}>1$. Here, $\xi_0=\cos^{-1}({\mit\Omega})$, and
\begin{equation}
A_+({\mit\Omega},\xi)= \frac{1}{2}\left[A(+1,{\mit\Omega},\xi) + A(-1,{\mit\Omega},\xi)\right].
\end{equation}

\subsection{Wide Island Limit}
In the so-called {\em wide island limit}, in which parallel electron heat transport dominates perpendicular heat transport,\cite{ntm2,island}
the electron temperature in the vicinity of the island chain can be written
\begin{equation}
T_e(X,\zeta) = T_{e\,l} + \varsigma\,W\,T_{e\,l}'\,\tilde{T}({\mit\Omega}),
\end{equation}
where $T_{e\,l}= T_{e\,0}(r_l)$ and $T_{e\,l}' = dT_{e\,0}(r_l)/dr$ are the equilibrium electron temperature and  temperature
gradient, respectively, at the island rational surface. 
Here, 
$\tilde{T}({\mit\Omega})$ satisfies\,\cite{ntm2}
\begin{equation}\label{e30}
\left\langle \left.\frac{\partial ^2\tilde{T}}{\partial X^2}\right|_\zeta \right\rangle =0,
\end{equation}
subject to the boundary condition that
\begin{equation}
\tilde{T}({\mit\Omega})\rightarrow |X|
\end{equation}
as $|X|\rightarrow \infty$. It
follows  from Eqs.~(\ref{eeven}), (\ref{e49}), and (\ref{e55})  that
\begin{equation}\label{e34}
\frac{d}{d{\mit\Omega}}\!\left(\langle Y^2\rangle\,\frac{d\tilde{T}}{d{\mit\Omega}}\right)=0
\end{equation}
subject to the boundary condition that
\begin{equation}
\tilde{T}({\mit\Omega})\rightarrow \frac{{\mit\Omega}^{1/2}}{\sqrt{8}}
\end{equation}
as ${\mit\Omega}\rightarrow\infty$. Note that $\tilde{T}({\mit\Omega})=0$ inside the magnetic separatrix, by symmetry, which implies that the electron
temperature profile is completely flattened in the region enclosed  by the separatix. \cite{ntm2}

Outside the separatrix,
\begin{equation}
\langle Y^2\rangle({\mit\Omega}) = \frac{1}{16}\int_0^{2\pi}\sigma(\xi)\sqrt{2\,({\mit\Omega}-\cos\xi)}\,\frac{d\xi}{2\pi}.
\end{equation}
Let 
\begin{equation}\label{kappa}
\kappa= \left(\frac{1+{\mit\Omega}}{2}\right)^{1/2}.
\end{equation}
Thus, the island O-points correspond to $\kappa=0$, and the magnetic separatrix to $\kappa=1$. 
It follows that 
\begin{equation}
\langle Y^2\rangle(\kappa) = \frac{\kappa}{4\pi}\int_0^{\pi/2}\sigma(2\,\vartheta-\pi)\left(1-\frac{\sin^2\vartheta}{\kappa^2}\right)^{1/2}\,d\vartheta
\end{equation}
for $\kappa>1$. 
Thus, making use of Eq.~(\ref{sigma}), 
\begin{equation}
\langle Y^2\rangle(\kappa) = \frac{\kappa}{4\pi}\,G(1/\kappa),
\end{equation}
where
\begin{align}
G(p) &=E_0(p) +2\sum_{\mu=1,\infty}\cos(\mu\,\pi)\,J_\mu(\mu\,\delta^2)\,E_\mu(p),\\[0.5ex]
E_\mu(p) &= \int_0^{\pi/2} \cos(2\,\mu\,\vartheta)\,(1-p^2\,\sin^2\vartheta)^{1/2}\,d\vartheta.
\end{align}

Equation~(\ref{e34}) yields
\begin{equation}
\tilde{T}(\kappa) = 0
\end{equation}
for $0\leq\kappa\leq 1$, and 
\begin{equation}
\frac{d}{d\kappa}\!\left[G(1/\kappa)\,\frac{d\tilde{T}}{d\kappa}\right]=0
\end{equation}
for $\kappa>1$. Thus,
\begin{equation}
\frac{d\tilde{T}}{d\kappa} = \frac{c}{G(1/\kappa)}
\end{equation}
for $\kappa>1$, subject to the boundary condition that
\begin{equation}
\tilde{T}(\kappa)\rightarrow \frac{\kappa}{2}
\end{equation}
as $\kappa\rightarrow \infty$. Now, $E_0(0) = \pi/2$, and  $E_{\mu>0}(0) = 0$,
which implies that $c=\pi/4$. So
\begin{align}\label{e47x}
\frac{d\tilde{T}}{d\kappa} &= \frac{\pi}{4}\,\frac{1}{G(1/\kappa)},\\[0.5ex]
\tilde{T}(\kappa) &= F(\kappa),\label{e48}\\[0.5ex]
F(\kappa) &= \frac{\pi}{4}\int_1^\kappa\frac{d\kappa'}{G(1/\kappa')}
\end{align}
for $\kappa>1$. 

\subsection{Helical Harmonics of Perturbed Electron Temperature}
We can write
\begin{equation}
\tilde{T}(X,\zeta)=\sum_{\nu=0,\infty}\delta T_\nu(X)\,\cos(\nu\,\zeta).
\end{equation}
Now,
\begin{equation}
\delta T_0(X) = \oint \tilde{T}(X,\zeta)\,\frac{d\zeta}{2\pi},
\end{equation}
where the integral is performed at constant $X$. It follows from Eqs.~(\ref{eeven}), (\ref{sigma}),  (\ref{kappa}), and (\ref{e48}) that
\begin{equation}
\delta T_0(X) = \int_0^{\xi_c}F(\kappa)\,\sigma(\xi)\,\frac{d\xi}{\pi},
\end{equation}
where 
\begin{equation}
\xi_c = \cos^{-1}(1-8\,Y^2)
\end{equation}
for $|Y|<1/2$, and $\xi_c=\pi$ for $|Y|\geq 1/2$. Furthermore,
\begin{equation}
\kappa =\left[4\,Y^2 +\cos^2\left(\frac{\xi}{2}\right)\right]^{1/2}.
\end{equation}

\begin{figure}
\centerline{\includegraphics[width=\textwidth]{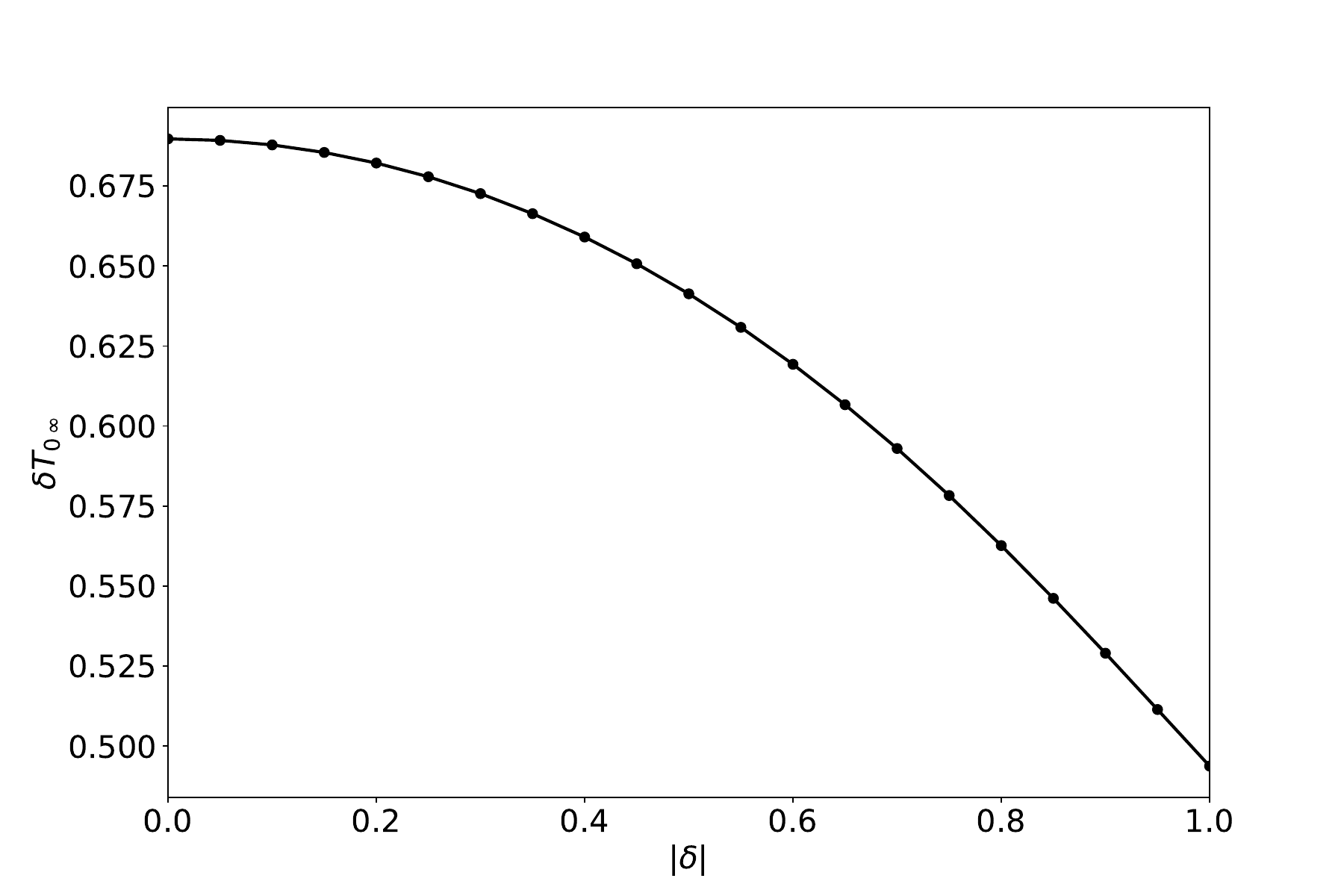}}
\caption{The island temperature flattening parameter, $T_{0\,\infty}$, plotted as a function of the modulus of the island asymmetry parameter, $|\delta|$.  \label{fig3}}
\end{figure}

Let 
\begin{align}\label{e79}
\delta T_{0\,+} &=\lim_{X\rightarrow \infty}\left[X - \delta T_0(X)\right],\\[0.5ex]
\delta T_{0\,-} &=-\lim_{X\rightarrow -\infty}\left[X - \delta T_0(X)\right],\\[0.5ex]
\delta T_{0\,\infty} &= \delta T_{0\,+}+ \delta T_{0\,-}.
\end{align}
The quantity $\delta T_{0\,\infty}$ is related to the reduction of the electron temperature in the plasma core, $\delta T_{e\,0}$, due to the flattening of the
temperature profile inside the island separatrix, as follows:
\begin{equation}\label{core}
\delta T_{e\,0} = W\,T_{e\,l}'\,\delta T_{0\,\infty}.
\end{equation}
Here, we are assuming that the equilibrium electron temperature at the plasma boundary is fixed.\cite{chang}
Figure~\ref{fig3} shows $T_{0\,\infty}$ plotted as a function of the modulus of the island asymmetry parameter, $|\delta|$.  
Note that $T_{0\,\infty}$ is positive, indicating that a magnetic island chain decreases the core electron temperature, assuming that the
unperturbed electron temperature gradient at the island rational surface is negative. [See Eq.~(\ref{core}).] It is clear that a symmetric (i.e., $\delta=0$) magnetic
island chain give rise to slightly larger reduction in the core temperature than an asymmetric island chain of the same width. 

For $\nu>0$, we have
\begin{equation}
\delta T_\nu(X) = 2\oint\tilde{T}(X,\zeta)\,\cos(\nu\,\zeta)\,\frac{d\zeta}{2\pi},
\end{equation}
where the integral is performed at constant $X$. 
Integrating by parts, we obtain
\begin{equation}
\delta T_\nu(X) = -\frac{2}{\nu}\oint\left.\frac{\partial \tilde{T}}{\partial\zeta}\right|_X\,\sin(\nu\,\zeta)\,\frac{d\zeta}{2\pi}.
\end{equation}
But,
\begin{equation}
\left.\frac{\partial \tilde{T}}{\partial\zeta}\right|_X=\frac{d\tilde{T}}{d\kappa}\left.\frac{\partial \kappa}{\partial\zeta}\right|_{X}
=\frac{1}{4\,\kappa}\frac{d\tilde{T}}{d\kappa}\,\left.\frac{\partial {\mit\Omega}}{\partial\zeta}\right|_{X}=-\frac{1}{4\,\kappa}\,\frac{d\tilde{T}}{d\kappa}\,\tau(\xi),
\end{equation}
where
\begin{equation}
\tau(\xi) = \sin\xi\,(1-\delta^2\,\cos\zeta)  -2\sqrt{8}\,\delta\,X\,\sin\zeta +\delta^2\,\sin(2\,\zeta),
\end{equation}
and use has been made of Eqs.~(\ref{e45}) and (\ref{kappa}). 
Hence,
\begin{equation}
\delta T_\nu(X) =\frac{1}{8\,\nu}\int_0^{\xi_c}\frac{\sin(\nu\,\zeta)\,\tau(\xi)\,\sigma(\xi)}{\kappa\,G(1/\kappa)}\,d\xi,
\end{equation}
where use has been made of Eqs.~(\ref{sigma}) and (\ref{e47x}).

\begin{figure}
\centerline{\includegraphics[width=\textwidth]{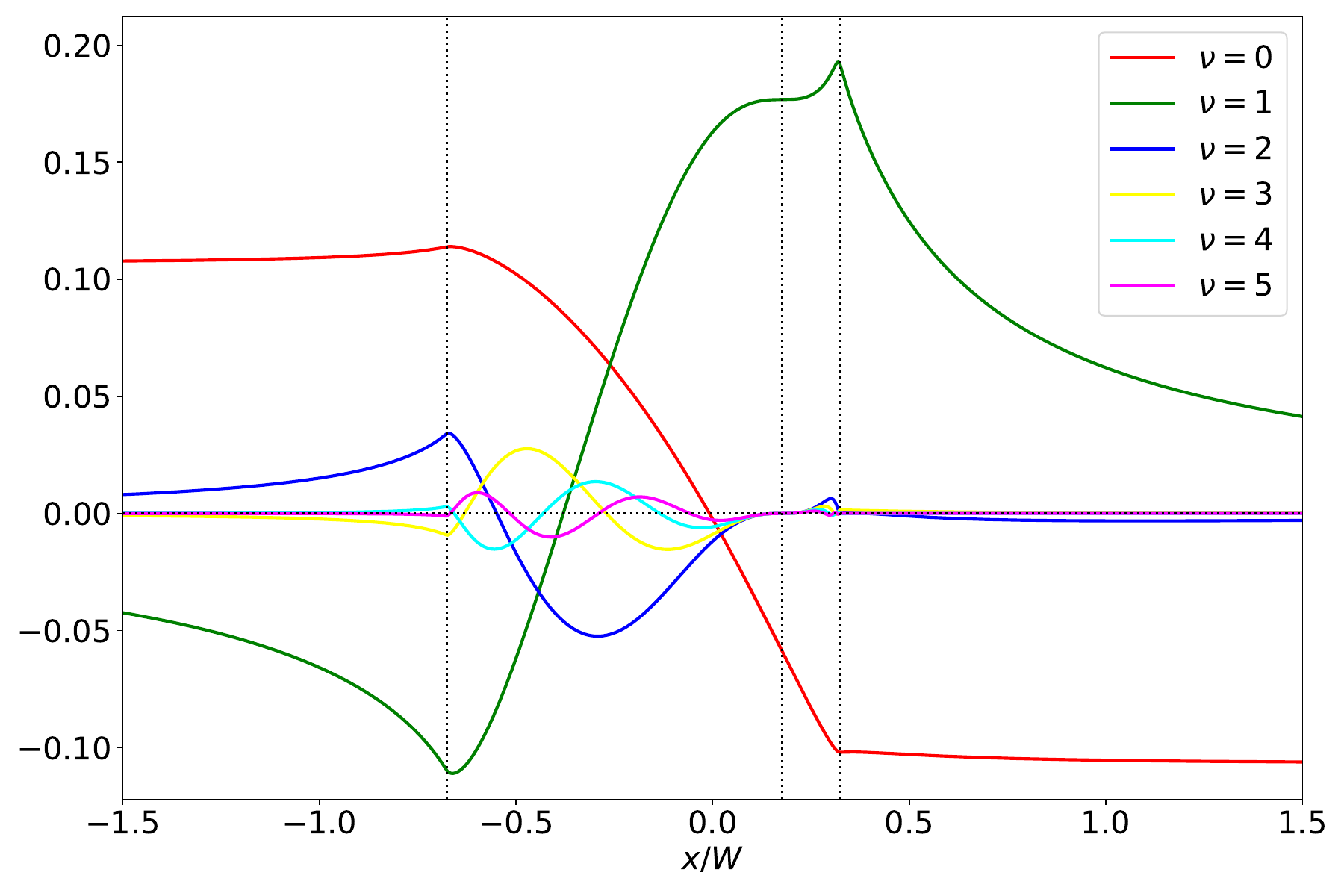}}
\caption{The helical harmonics of the normalized electron temperature  in the inner region, $\delta T_\nu(x/W)$, calculated for an asymmetric
magnetic island characterized by $\delta=0.5$.  The curve labelled $0$ actually shows $[\delta T_0(x/W)-x/W]/3$, whereas the
curve labelled  $1$ actually
shows $\delta T_1(x/W) +  \delta/\sqrt{8}$. The vertical dotted lines show the locations of the inner limit of the magnetic separatrix,
the island X-point, and the outer limit of the magnetic separatrix, in order from the left to the right.\label{fig4}}
\end{figure}

Figure~\ref{fig4} shows the harmonics of the normalized electron temperature  in the inner region, $\delta T_\nu(x/W)$, calculated for an asymmetric
magnetic island characterized by $\delta=0.5$.  Note that the harmonics are asymmetric in $X$. (By contrast, Fig.~3 of Ref.~\onlinecite{ntm2} shows
the purely anti-symmetric harmonics of a symmetric island.) It can be seen that the $\nu=0$ and $\nu=1$ harmonics extend into the outer region, whereas the
$\nu>1$ harmonics are strongly localized in the vicinity of the island. 

\begin{figure}
\centerline{\includegraphics[width=\textwidth]{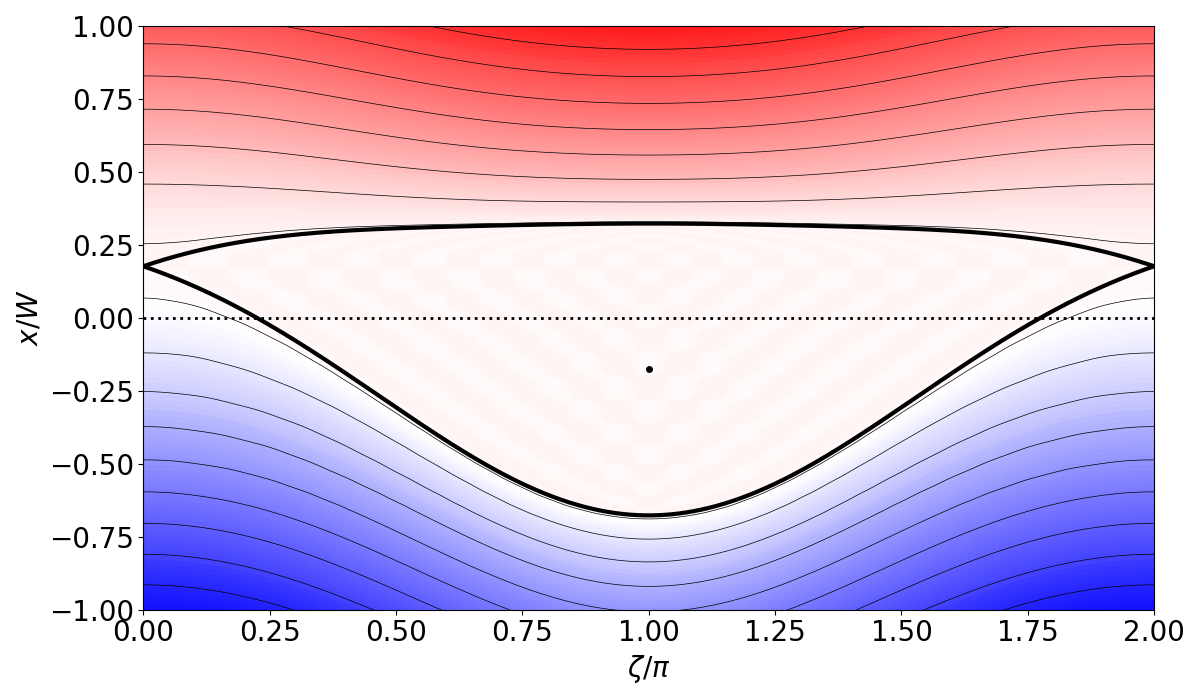}}
\caption{Contours of the normalized electron temperature profile, $\tilde{T}(x/W,\zeta)$, in the vicinity of an asymmetric magnetic island  characterized by
$\delta = 0.5$. The thick solid line shows the magnetic separatrix, the dotted line shows the rational surface, and the black dot shows the
island O-point.\label{fig5}}
\end{figure}

Finally, Fig.~\ref{fig5} shows the normalized electron temperature distribution, $\tilde{T}(x/W,\zeta)$, in the vicinity of an asymmetric magnetic island  characterized by
$\delta = 0.5$. This temperature distribution
is reconstructed from 16 helical harmonics (i.e., $\nu$ in the range 0 to 15). As expected, the temperature profile is almost completely flattened in the region enclosed  within the magnetic separatrix. 

\subsection{Modified Rutherford Equation}
The nonlinear growth of the magnetic island  chain associated with an NTM that is resonant at the $l$th rational surface is governed by a modified Rutherford equation that takes the form\,\cite{ntm1,ntm4,island,boot,fitz}
\begin{equation}\label{eruth}
G_{\rm ruth}\,\tau_R\,\frac{d}{dt}\!\left(\frac{W}{r_l}\right) = E_{ll} +\left[\hat{\beta}\,(\alpha_b-\alpha_c)\,G_{\rm boot} + J_{\rm max}\,G_{\rm eccd}\right]
\frac{r_l}{W},
\end{equation}
where
\begin{align}
G_{\rm ruth} &= 2\int_{-1}^\infty \frac{(\langle\cos\xi\rangle + \delta^2\,\langle \sin\xi\,\sin\zeta\rangle)\,\langle \cos\zeta\rangle}{\langle 1\rangle}\,d{\mit\Omega},\\[0.5ex]
G_{\rm boot} &= \int_1^\infty \frac{\langle \cos\zeta\rangle}{\langle 1\rangle\,\langle Y^2\rangle}\,d{\mit\Omega},\\[0.5ex]
G_{\rm eccd} &= - 16\int_{-1}^\infty \frac{\langle J_+\rangle\,\langle \cos\zeta\rangle}{\langle 1\rangle}\,d{\mit\Omega},\label{geccd}
\end{align}
Here,  $J_+(x,\zeta)$ is the component of the normalized  (such that the peak value is unity) current density profile driven by electron cyclotron waves that is even in $Y$. 
Moreover, 
\begin{align}
\tau_R &= \left(\frac{\mu_0\,r^2\,R_0^{\,2}}{\eta_\parallel}\right)_{r_l},\\[0.5ex]
\hat{\beta} &= \left(\frac{\mu_0\,n_e\,T_{e\,0}}{B_0^{\,2}}\right)_{r_l}\frac{L_s}{L_T}\,\\[0.5ex]
L_s&=\left(\frac{q}{s}\right)_{r_l},\\[0.5ex]
L_T &= -\left(\frac{T_{e\,0}}{dT_{e\,0}/dr}\right)_{r_l},\\[0.5ex]
\alpha_b &=(\beta_{11}-\beta_{12})\, \left(f_t\,\frac{q}{r}\right)_{r_l},\\[0.5ex]
f_t&= 1.46\,r^{1/2},\\[0.5ex]
\alpha_c &= \frac{2\,L_s}{L_c},\\[0.5ex]
L_c &= \left[\frac{1}{r\,(1-1/q^2) -a\,s\,H_1'}\right]_{r_l}.
\end{align}
Here, $E_{ll}$ is the  normalized linear tearing stability index of an $m_l$, $n$ tearing mode
 that only reconnects magnetic flux at the $l$th rational surface,\cite{tear5} $\tau_R$ is the resistive diffusion timescale, $\eta_\parallel(r)$  the plasma parallel electrical resistivity, $\hat{\beta}$ the normalized electron pressure, $L_s$  the (normalized) magnetic shear-length, 
$L_T$ the (normalized) electron temperature gradient scale-length, $L_c$ the (normalized) average magnetic field-line curvature scale-length, and $f_t$  the fraction of trapped particles. Moreover, for a plasma with an effective charge
number of unity, $\beta_{11}=1.641$ and $\beta_{12}= 1.225$.\cite{fitz}
Finally,
\begin{equation}\label{ecur}
J_{\rm max} = \frac{\mu_0\,R_0\,L_s}{B_0}\,j_{\rm max},
\end{equation}
where $j_{\rm max}$ is the unnormalized peak current density driven by electron cyclotron waves.  
 
 The term in the modified Rutherford equation, (\ref{eruth}),  that involves $\alpha_b$  represents  the
 destabilizing effect of the loss of the bootstrap current inside the island separatrix consequent on the flattening of the electron temperature profile. \cite{ntm2,car} The term involving $\alpha_c$ represents  the stabilizing effect of  magnetic field-line curvature  consequent on the flattening of the electron temperature profile.\cite{fitz,kot} Finally, the term involving $J_{\rm max}$ represents the effect of current driven in the island region by electron cyclotron waves.\cite{island} In writing the modified
 Rutherford equation, we have adopted various large aspect-ratio approximations,\cite{ggj1,fitz} have assumed that the driven current rapidly equilibrates on
 island magnetic flux-surfaces, 
and have  neglected the contributions to the equation due to incomplete temperature flattening\,\cite{ntm2} and the ion polarization current\,\cite{polz} (which are only important for
 very narrow islands). We  have also neglected the contribution due to
 plasma heating via electron cyclotron waves (which is similar to, but generally smaller than, that of electron cyclotron current drive).\cite{ntm4,island}
 
 The following results are useful when performing flux-surface averages:\,\cite{island}
\begin{equation}
\langle A(\kappa,\xi)\rangle(\kappa) = \frac{1}{\pi}\int_0^{\pi/2}\frac{\sigma(\xi)\,A(\kappa,\xi)}{\sqrt{1-\kappa^2\,\sin^2\vartheta}}\,d\vartheta
\end{equation}
for $0\leq \kappa\leq 1$, where $\xi=2\,\cos^{-1}(\kappa\,\sin\vartheta)$. Likewise, 
\begin{equation}
\langle A(\kappa,\xi)\rangle(\kappa) = \frac{1}{\pi}\int_0^{\pi/2}\frac{\sigma(\xi)\,A(\kappa,\xi)}{\sqrt{\kappa^2-\sin^2\vartheta}}\,d\vartheta
\end{equation}
for $\kappa>1$, where $\xi=\pi-2\,\vartheta$. Recall that $\kappa=[(1+{\mit\Omega})/2]^{1/2}$. 

Table~\ref{t1} lists various quantities calculated by the TJ code at the $3$, $2$ and the $2$, $1$ rational surfaces for the example plasma equilibrium shown in Figs.~\ref{figa} and \ref{figb}. 
Note that the critical linear tearing stability index that must be exceeded before the stabilizing effect of average magnetic field-line curvature is overcome\,\cite{tear10} exceeds the linear tearing stability index for both surfaces. In other words, the $3$, $2$ and the $2$, $1$ classical tearing modes are both linearly stable. On the other hand, the difference between the bootstrap parameter, $\alpha_b$, and the curvature parameter, $\alpha_c$, is positive
for both surfaces. In other words, the $3$, $2$ and the $2$, $1$ neoclassical tearing modes  are both potentially unstable. 

Figure~\ref{fig6} shows the integrals $G_{\rm ruth}$ and $G_{\rm boot}$ evaluated as functions of the modulus of the island asymmetry
parameter, $|\delta|$. It can be seen that both integrals only depend weakly on $|\delta|$, as long as $|\delta|$ does not get too close to
unity.\cite{ece6,island} Note that $G_{\rm ruth} = 0.8360$ and $G_{\rm boot} =6.381$ for an island whose asymmetry parameter is $0.2$. 
 
\subsection{ECCD Deposition Profile}
 Let us assume that the normalized profile of the current density driven by electron cyclotron waves in the island region (prior
 to equilibration around flux-surfaces) takes the form
 \begin{equation}
 J(x,\zeta) = \exp\left[-\frac{(x-d)^2}{2\,D^{\,2}}\right]\left[\frac{1+\cos(\zeta-{\mit\Delta}\zeta)}{2}\right],
 \end{equation}
 where $D$ is the radial width of the profile, $d$  the radial offset between the peak current and the rational surface, and
${\mit\Delta}\zeta$ the angular offset between the peak current and the island O-point. Note that the profile is  comparatively narrow in $x$, and comparatively
wide in $\zeta$, as is  generally the case in experiments. It turns out that  only the component of $J(x,\zeta)$ that is even in $\zeta$ contributes to the integral (\ref{geccd}), so we can effectively write
\begin{equation}
J(x,\zeta)= \exp\!\left[-\frac{(x-d)^2}{2\,D^{\,2}}\right]\left(\frac{1+\cos\zeta\,\cos{\mit\Delta}\zeta}{2}\right).
\end{equation}
Let $\hat{D}=D/W$ and $\hat{d}=d/W$. Making use of Eq.~(\ref{ek}), we obtain 
\begin{equation}
J(\varsigma,Y,\zeta) =  \exp\!\left[-\frac{(\varsigma\,Y+\delta\,\cos\zeta/\sqrt{8}-\hat{d})^2}{2\,\hat{D}^{\,2}}\right]\left(\frac{1+\cos\zeta\,\cos{\mit\Delta}\zeta}{2}\right),
\end{equation}

Let 
\begin{align}
J_O(\varsigma,Y,\zeta) &=  \exp\!\left[-\frac{(\varsigma\,Y+\delta\,\cos\zeta/\sqrt{8}-\hat{d})^2}{2\,\hat{D}^{\,2}}\right]\left(\frac{1+\cos\zeta}{2}\right),\\[0.5ex]
J_X(\varsigma,Y,\zeta) &=  \exp\!\left[-\frac{(\varsigma\,Y+\delta\,\cos\zeta/\sqrt{8}-\hat{d})^2}{2\,\hat{D}^{\,2}}\right]\left(\frac{1-\cos\zeta}{2}\right),\\[0.5ex]
J_{O+}(Y,\zeta) &= \frac{J_O(1,Y,\zeta) + J_O(-1,Y,\zeta)}{2},\\[0.5ex]
J_{X+}(Y,\zeta) &= \frac{J_X(1,Y,\zeta) + J_X(-1,Y,\zeta)}{2}.
\end{align}
It follows from Eq.~(\ref{geccd}) that
\begin{align}
G_{\rm eccd}({\mit\Delta}\zeta)& = G_{{\rm eccd}\,O}\left(\frac{1+\cos{\mit\Delta}\zeta}{2}\right)+G_{{\rm eccd}\,X}\left(\frac{1-\cos{\mit\Delta}\zeta}{2}\right),\\[0.5ex]
G_{{\rm eccd}\,O} &= - 16\int_{-1}^\infty \frac{\langle J_{O+}\rangle\,\langle \cos\zeta\rangle}{\langle 1\rangle}\,d{\mit\Omega},\\[0.5ex]
G_{{\rm eccd}\,X} &= - 16\int_{-1}^\infty \frac{\langle J_{X+}\rangle\,\langle \cos\zeta\rangle}{\langle 1\rangle}\,d{\mit\Omega}.
\end{align}
Note that if there were no peaking of the current density profiles driven by electron cyclotron waves in the angular variable $\zeta$ (i.e., if the profile were
independent of $\zeta$) then the integral $G_{\rm eccd}$ would take the value  $2\,G_{\rm eccd}({\mit\Delta}\zeta= \pi/2)$.

\subsection{Results}
\begin{figure}
\centerline{\includegraphics[width=\textwidth]{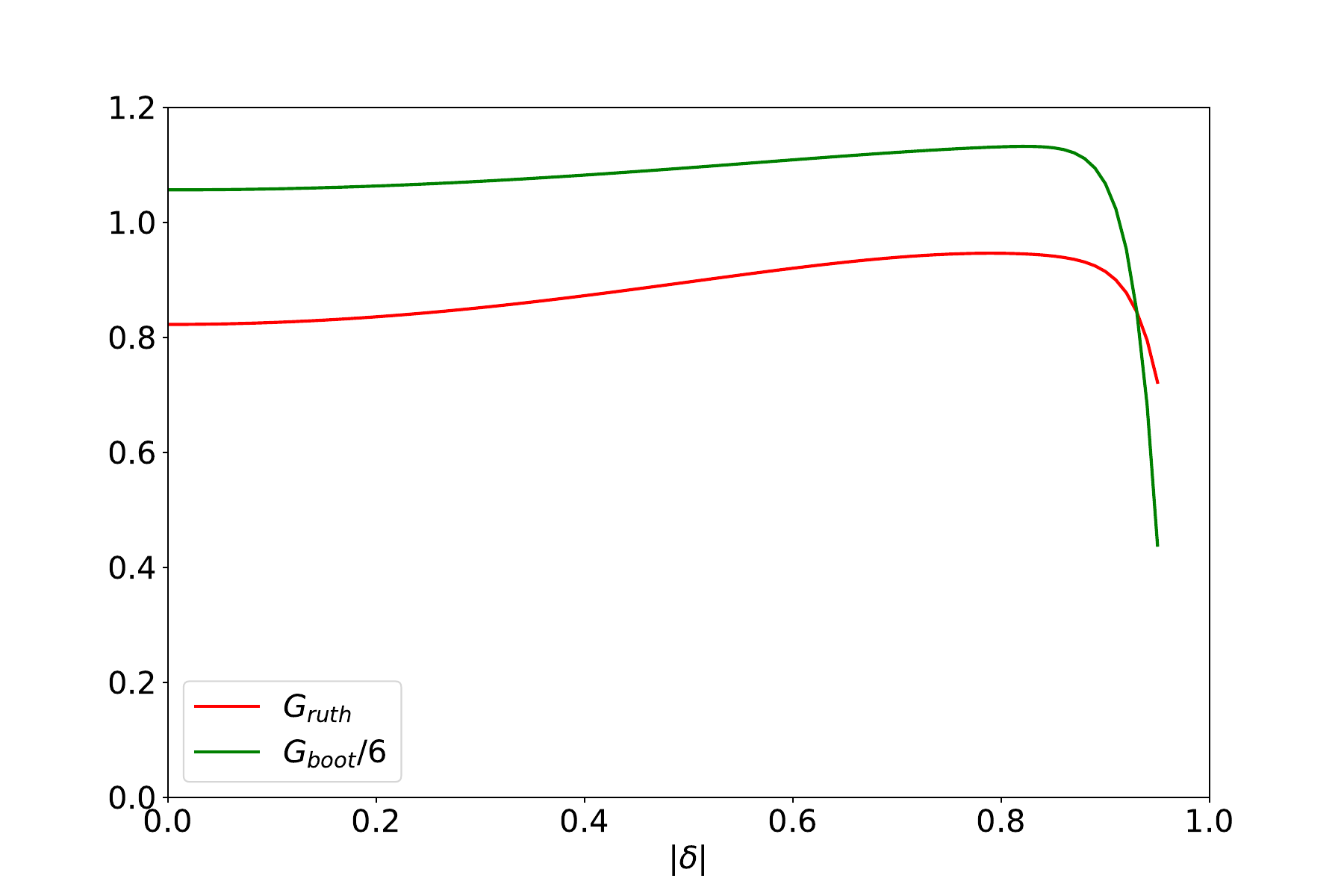}}
\caption{The integrals $G_{\rm ruth}$ and $G_{\rm boot}/6$ evaluated as functions of the modulus of the island asymmetry parameter, $|\delta|$.  \label{fig6}}
\end{figure}

\begin{figure}
\centerline{\includegraphics[width=\textwidth]{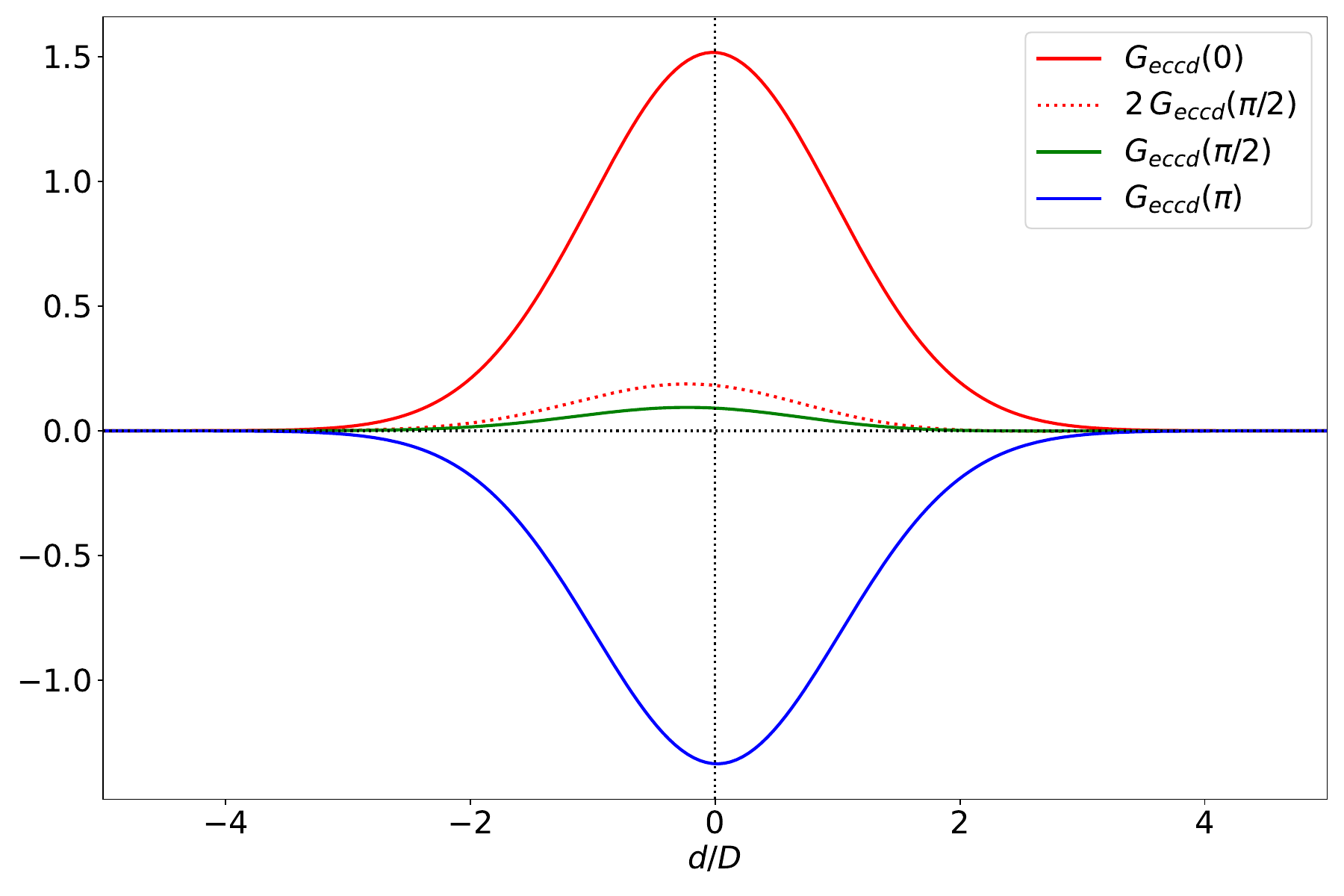}}
\caption{The integral $G_{\rm eccd}({\mit\Delta}\zeta)$ evaluated as a function of $d/D$ at ${\mit\Delta}\zeta = 0$, $\pi/2,$ and $\pi$ for an  island
of full width $W=0.1\,D$ and asymmetry parameter $\delta=0.5$. The dotted curve shows the integral when the driven current profile is independent of $\zeta$. \label{fig7}}
\end{figure}

Figure~\ref{fig7} shows the integral $G_{\rm eccd}$ evaluated as a function of $d/D$ for a thin island of width $W=0.1\,D$ and asymmetry parameter
$\delta=0.5$. Now, it is clear from the modified Rutherford equation, (\ref{eruth}), as well as from  Fig.~\ref{fig6}, that in order for ECCD to suppress an NTM to
such an extent that the island width falls below the threshold value needed to trigger the mode, implying that the mode is completely
stabilized, the integral $G_{\rm eccd}$ needs to be finite and positive in the limit as $W/D\rightarrow 0$. It is apparent  from Fig.~\ref{fig6} that this is the case
as long as $|d|\lesssim 2\,D$ and ${\mit\Delta}\zeta\lesssim \pi/2$. In other words, successful stabilization is possible provided  the radial
offset of the peak current driven by electron cyclotron waves from the rational surface does not exceed twice the radial standard deviation of the current drive deposition profile, 
and as long as the angular offset of the peak current from the island O-point does not exceed $\pi/2$. The figure also indicates that for thin islands
there is a considerable benefit to be had from peaking the deposition profile in $\zeta$ in the vicinity of the O-point (which, in practice, is achieved by
modulating the electron cyclotron source such that it is only turned on the when the island O-point is directly in the line of fire), rather than having the profile independent of
$\zeta$ (which, in practice, is achieved by not modulating the source).\cite{ece6} (Here, we are assuming that the island chain is rotating, as is generally the case in experiments.) 

Note that if the ECCD is optimal
(i.e., $d=0$ and ${\mit\Delta}\zeta=0$) then $G_{\rm eccd}$ attains a maximum value of $1.433$ for an island whose asymmetry parameter is $\delta=0.2$. 
Making use of Eqs.~(\ref{eruth}) and (\ref{ecur}),  as well as the information in Table~\ref{t1} and Fig.~\ref{fig6}, 
we deduce that the critical normalized peak driven current density at the $q=3/2$ surface needed to stabilize a 3, 2 NTM with an asymmetry parameter of $\delta = 0.2$ is $J_{\rm max\,crit}= 0.48$, which corresponds
to an unnormalized peak current density of $j_{\rm max\,crit}= 1.9\times 10^5\,{\rm A/m^2}$. Likewise, the critical normalized peak driven current density at the $q=2$ surface needed to stabilize a 2, 1 NTM with an asymmetry parameter of $\delta = 0.2$ is $J_{\rm max\,crit}= 0.27$, which corresponds
to an unnormalized peak current density of $j_{\rm max\,crit}= 1.5\times 10^5\,{\rm A/m^2}$

\begin{figure}
\centerline{\includegraphics[width=\textwidth]{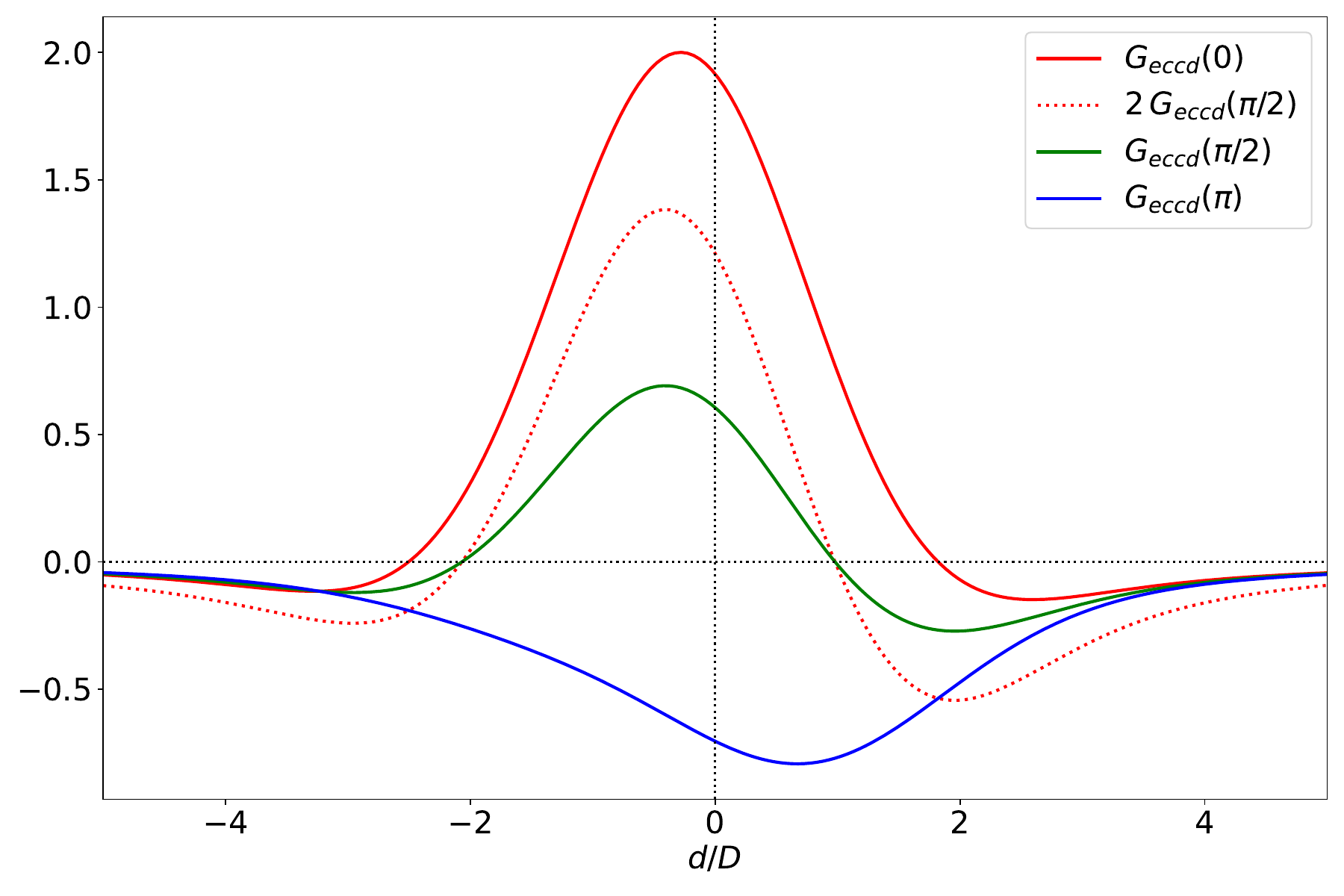}}
\caption{The integral $G_{\rm eccd}({\mit\Delta}\zeta)$ evaluated as a function of $d/D$ at ${\mit\Delta}\zeta = 0$, $\pi/2,$ and $\pi$  for an  island
of full width $W=2.0\,D$ and asymmetry parameter $\delta=0.5$. The dotted curve shows the integral when the driven current profile is independent of $\zeta$. \label{fig8}}
\end{figure}

\begin{figure}
\centerline{\includegraphics[width=\textwidth]{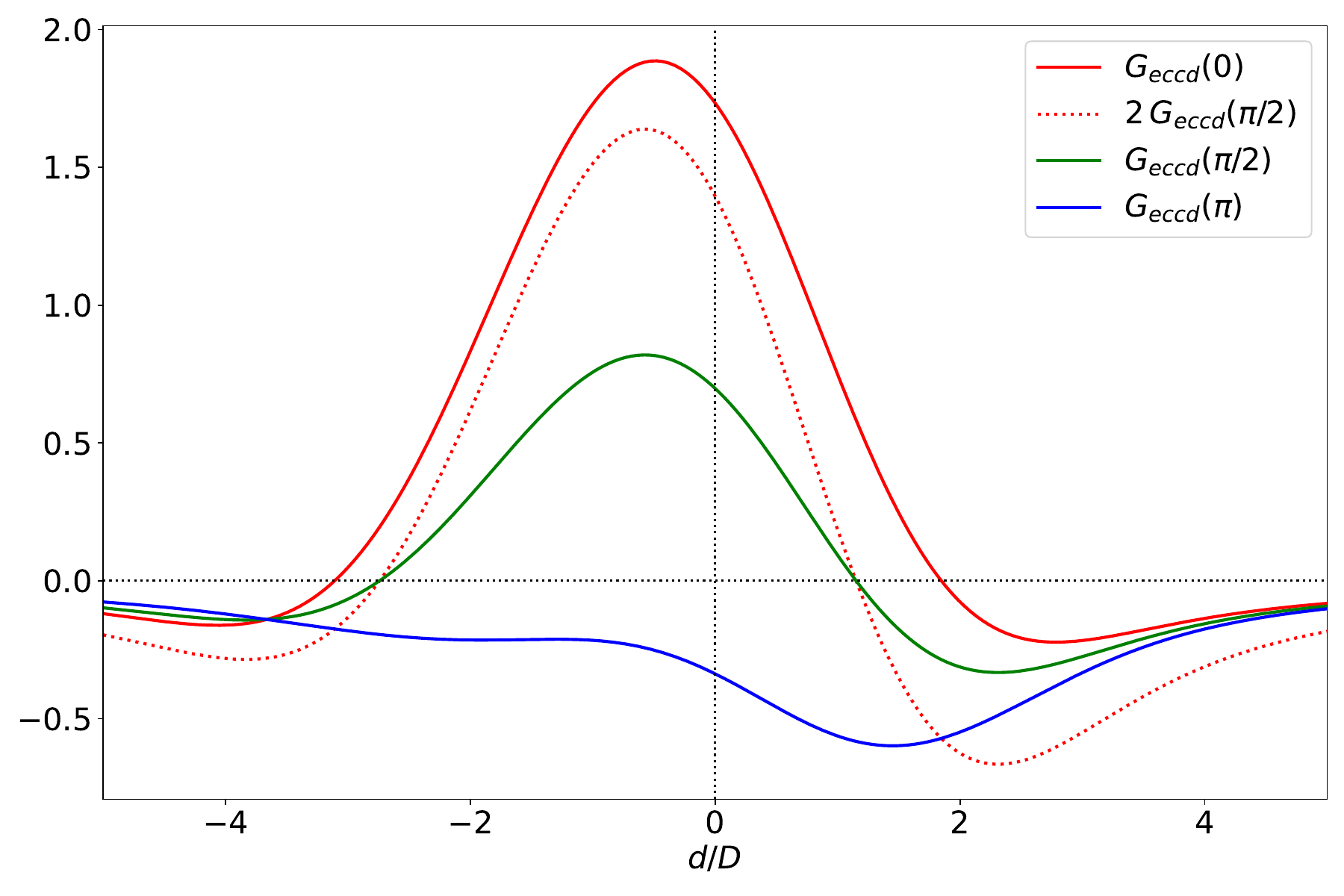}}
\caption{The integral $G_{\rm eccd}({\mit\Delta}\zeta)$ evaluated as a function of $d/D$ at ${\mit\Delta}\zeta = 0$, $\pi/2,$ and $\pi$ for an  island
of full width $W=4.0\,D$ and asymmetry parameter $\delta=0.5$. The dotted curve shows the integral when the driven current profile is independent of $\zeta$.\label{fig9}}
\end{figure}

Figures~\ref{fig8} and \ref{fig9} show the integral $G_{\rm eccd}$ evaluated  as a function of $d/D$ for wide islands of width $W=2\,D$ and $4\,D$, respectively, and asymmetry parameter
$\delta=0.5$. It can be seen that, as the island increases in width, the optimum radial location of the ECCD profile shifts inward from the rational
surface.\cite{ece6} This is indicative of the fact that the true target for ECCD is the island O-point (which is shifted inward from the
rational surface a distance $W\,\delta/\sqrt{8}$) rather than the rational surface. It is again the case that the radial offset of the peak current from the island
O-point needs to be less that twice the radial standard deviation of the current drive profile. 
Figures~\ref{fig8} and \ref{fig9}  also suggest that the benefit of modulating the
electron cyclotron source is considerably reduced for wide islands compared to narrow islands. 

\section{Global Perturbed Electron Temperature}\label{s5}
\subsection{Asymptotic Matching}
The asymptotic matching process consists of writing the NTM-modified electron temperature profile in the form
\begin{align}
T_e(r,\theta,\phi)&= T_{e\,0}(r) +\delta T_{e\,+}-{\mit\Psi}_+\frac{q(r)}{r\,g(r)}\,\frac{T_{e\,0}'(r)\,\hat{\psi}_{m_l}(r)}{m_l-n\,q(r)}\,{\rm e}^{\,{\rm i}\,(m_l\,\theta-n\,\phi)}\nonumber\\[0.5ex]&\phantom{=}
-{\mit\Psi}\sum_{j=1,J}^{m_j\neq m_l} \frac{q(r)}{r\,g(r)}\,\frac{T_{e\,0}'(r)\,\hat{\psi}_{m_j}(r)}{m_j-n\,q(r)}\,{\rm e}^{\,{\rm i}\,(m_j\,\theta-n\,\phi)}
\end{align}
in the segment of the outer region that lies outside the rational surface at which the NTM reconnects magnetic flux: i.e., $r>r_l+W$. Here, $T_{e\,0}'(r) = dT_{e\,0}/dr$. Likewise, the
island-modified electron temperature profile takes the form
\begin{align}
T_e(r,\theta,\phi)&= T_{e\,0}(r) +\delta T_{e\,-}-{\mit\Psi}_-\frac{q(r)}{r\,g(r)}\,\frac{T_{e\,0}'(r)\,\hat{\psi}_{m_l}(r)}{m_l-n\,q(r)}\,{\rm e}^{\,{\rm i}\,(m_l\,\theta-n\,\phi)}\nonumber\\[0.5ex]&\phantom{=}
-{\mit\Psi}\sum_{j=1,J}^{m_j\neq m_l} \frac{q(r)}{r\,g(r)}\,\frac{T_{e\,0}'(r)\,\hat{\psi}_{m_j}(r)}{m_j-n\,q(r)}\,{\rm e}^{\,{\rm i}\,(m_j\,\theta-n\,\phi)}
\end{align}
in the segment of the outer region that lies inside the rational surface: i.e., $r<r_l-W$. Finally, the NTM-modified electron temperature profile 
takes the form
\begin{align}
T_e(r,\theta,\phi) &= T_{e\,l} + T_{e\,l}'\,W\sum_{\nu=0,\nu_{\rm max}}\delta T_\nu(x/W)\,{\rm e}^{\,{\rm i}\,\nu\,(m_l\,\theta-n\,\phi)}\nonumber\\[0.5ex]
&\phantom{=}
-{\mit\Psi}\sum_{j=1,J}^{m_j\neq m_l} \frac{q(r)}{r\,g(r)}\,\frac{T_{e\,0}'(r)\,\hat{\psi}_{m_j}(r)}{m_j-n\,q(r)}\,{\rm e}^{\,{\rm i}\,(m_j\,\theta-n\,\phi)}
\end{align}
in the inner region: i.e., $r_l-W<r<r_l+W$.  Continuity of the solution at $r=r_l\pm W$ implies that
\begin{align}
\delta T_{e\,+} &= T_{e\,l}'\,W\,\delta T_{0}(1)+T_{e\,l}'\,W\,\delta T_{0\,+} - T_{e\,l}'\,W,\\[0.5ex]
\delta T_{e\,-} &= T_{e\,l}'\,W\,\delta T_{0}(-1) +T_{e\,l}'\,W\,\delta T_{0\,+}+ T_{e\,l}'\,W,\\[0.5ex]
{\mit\Psi}_{+} &= - T_{e\,l}'\,W\,\delta T_1(1)\left(\frac{r\,g}{q}\,\frac{m_l-n\,q}{T_{e\,0}'\,\hat{\psi}_{m_l}}\right)_{r_l+W},\\[0.5ex]
{\mit\Psi}_{l-} &=- T_{e\,l}'\,W\,\delta T_1(-1)\left(\frac{r\,g}{q}\,\frac{m_l-n\,q}{T_{e\,0}'\,\hat{\psi}_{m_l}}
\right)_{r_l-W}.
\end{align}
Here, the NTM island width, $W$, is specified, and the reconnected magnetic flux, ${\mit\Psi}$, and the island asymmetry parameter, $\delta$,  are then deduced from Eqs.~(\ref{e46}) and (\ref{e47}), respectively. Moreover,
$m_l$, $n$ are, respectively, the poloidal and toroidal mode numbers of the NTM, $r_l$ is the minor radius at which the NTM reconnects magnetic
flux, $T_{e\,l}=T_{e\,0}(r_l)$, $T_{e\,l}' = T_{e\,0}'(r_l)$, $\delta T_{0\,+}$ is defined in Eq.~(\ref{e79}), $\delta T_{e\,+}$ is the reduction in the equilibrium electron temperature profile outside the
island rational surface, and $\delta T_{e\,-}$ is the corresponding reduction inside the rational surface. The asymptotic matching process assumes that
the $\delta T_{\nu>1}$ are negligible at $r= r_l\pm W$. (See Fig.~\ref{fig4}.)

 Note that if we were to neglect the nonlinearly generated overtone harmonics of the $m_l$, $n$ harmonic in the inner
region then the temperature perturbation at a general toroidal angle, $\phi$, could be expressed as  a linear combination of the temperature perturbations at $\phi=0$ and
$\phi=\pi/(2\,n)$. However, the presence of the overtone harmonics spoils this scheme (because the harmonics have different toroidal mode numbers as
well as different poloidal mode numbers). Hence, it is necessary to separately calculate the temperature perturbation at a large number of equally-spaced values of $\phi$. 

\subsection{Results}
Table~\ref{t2} shows various parameters derived from the asymptotic matching process for a 3, 2 and a 2, 1 NTM in the example plasma equilibrium pictured in Figs.~\ref{figa} and
\ref{figb}. In both cases, we have chosen rather wide islands of width $W=0.1\,a$ for ease of visualization. The asymmetry parameter for the 3, 2 mode is $\delta=0.272$,
whereas that for the 2, 1 mode is $\delta=0.150$. This implies that both modes are characterized by radial asymmetric island chains in which the island O-points are
shifted radially inward from the rational surfaces. Both modes are also (by design) characterized by rather small reductions in the equilibrium electron temperature outside the
rational surface, and quite substantial reductions inside the rational surfaces.

\begin{figure}
\centerline{\includegraphics[width=\textwidth]{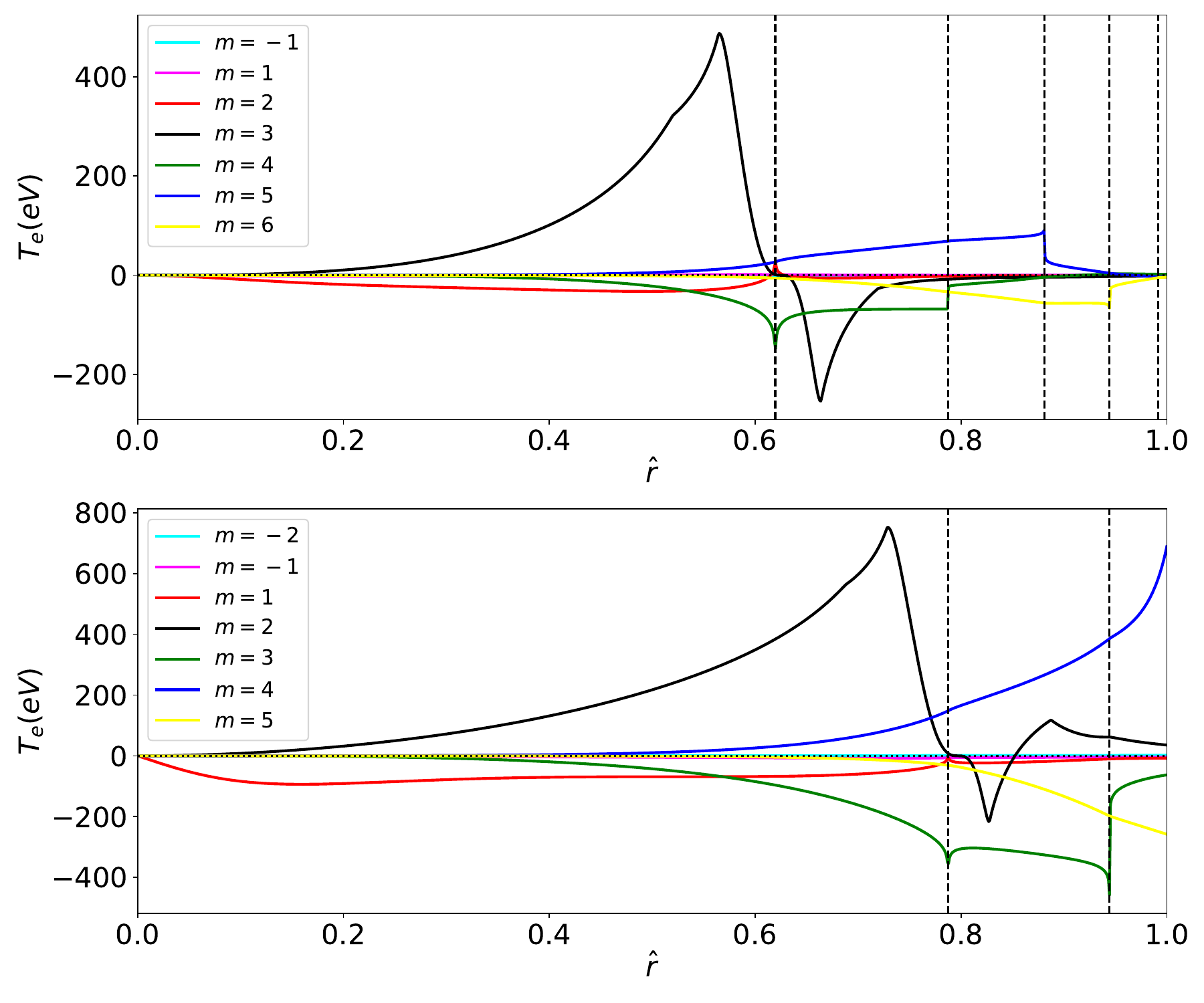}}
\caption{Harmonics of the perturbed electron temperature associated with a 3, 2 (top panel) and a 2, 1 (bottom panel) NTM of island width $W=0.1\,a$ in the example plasma equilibrium pictured in Figs.~\ref{figa}
and \ref{figb}. The harmonics all have the same toroidal mode number as the NTM. The vertical dashed lines show the positions of the rational surfaces. \label{fig10}}
\end{figure}

 Figure~\ref{fig10} shows the various harmonics of the 3, 2 and 2, 1 NTMs as functions of the flux-surface
label $r$. Note that
the overtone harmonics in the island regions are not shown in this figure. It can be seen that both NTMs consist of many coupled poloidal harmonics. 

\begin{figure}
\centerline{\includegraphics[width=\textwidth]{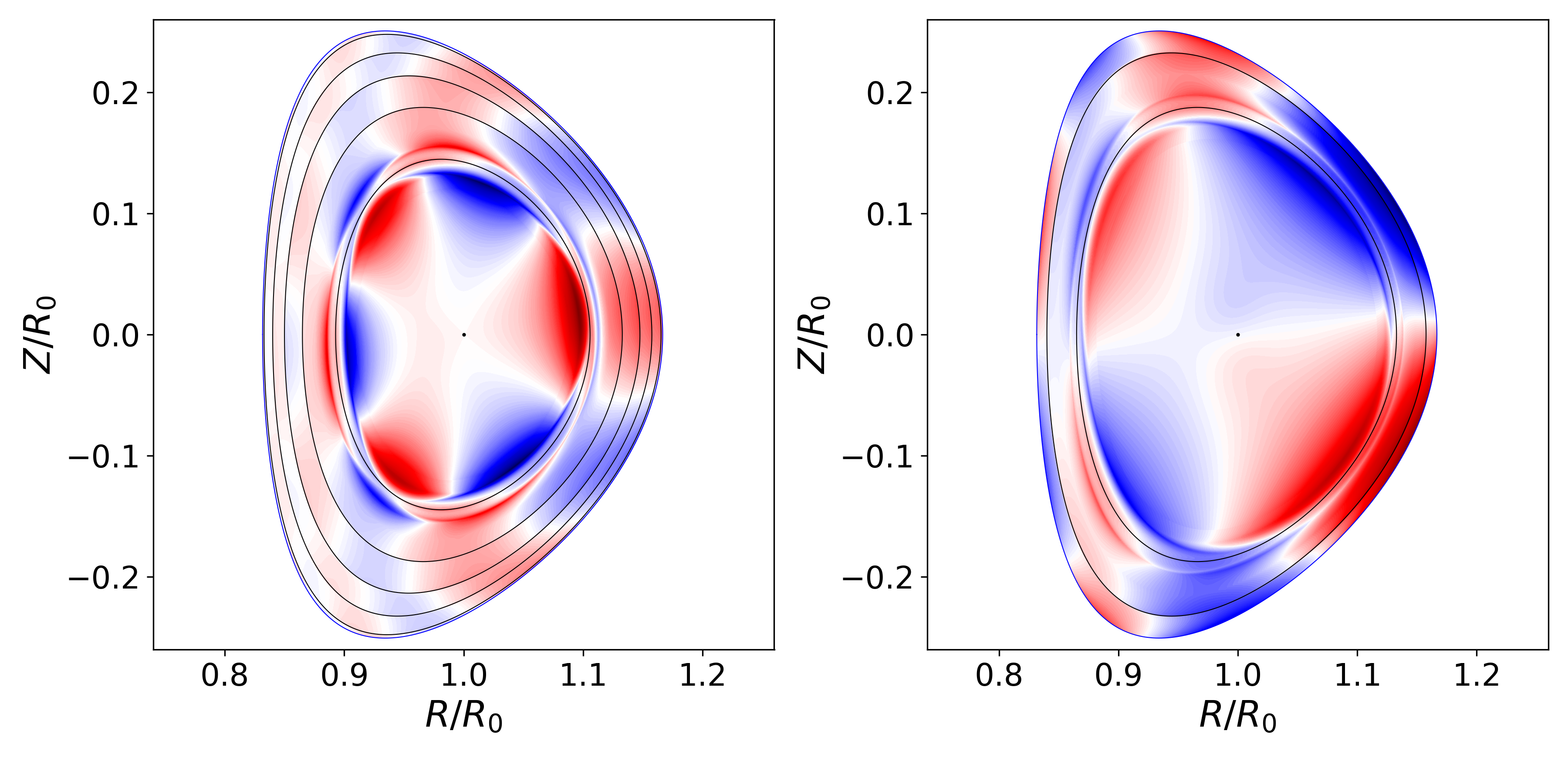}}
\caption{Electron temperature perturbation at a particular toroidal angle associated with a 3, 2 (left panel) and a 2, 1 (right panel) NTM of island width $W=0.1\,a$ in the example plasma equilibrium pictured in Figs.~\ref{figa}
and \ref{figb}. The black curves show the locations of the rational surfaces. The black dot shows the location of the magnetic axis.\label{fig11}}
\end{figure}

\begin{figure}
\centerline{\includegraphics[width=\textwidth]{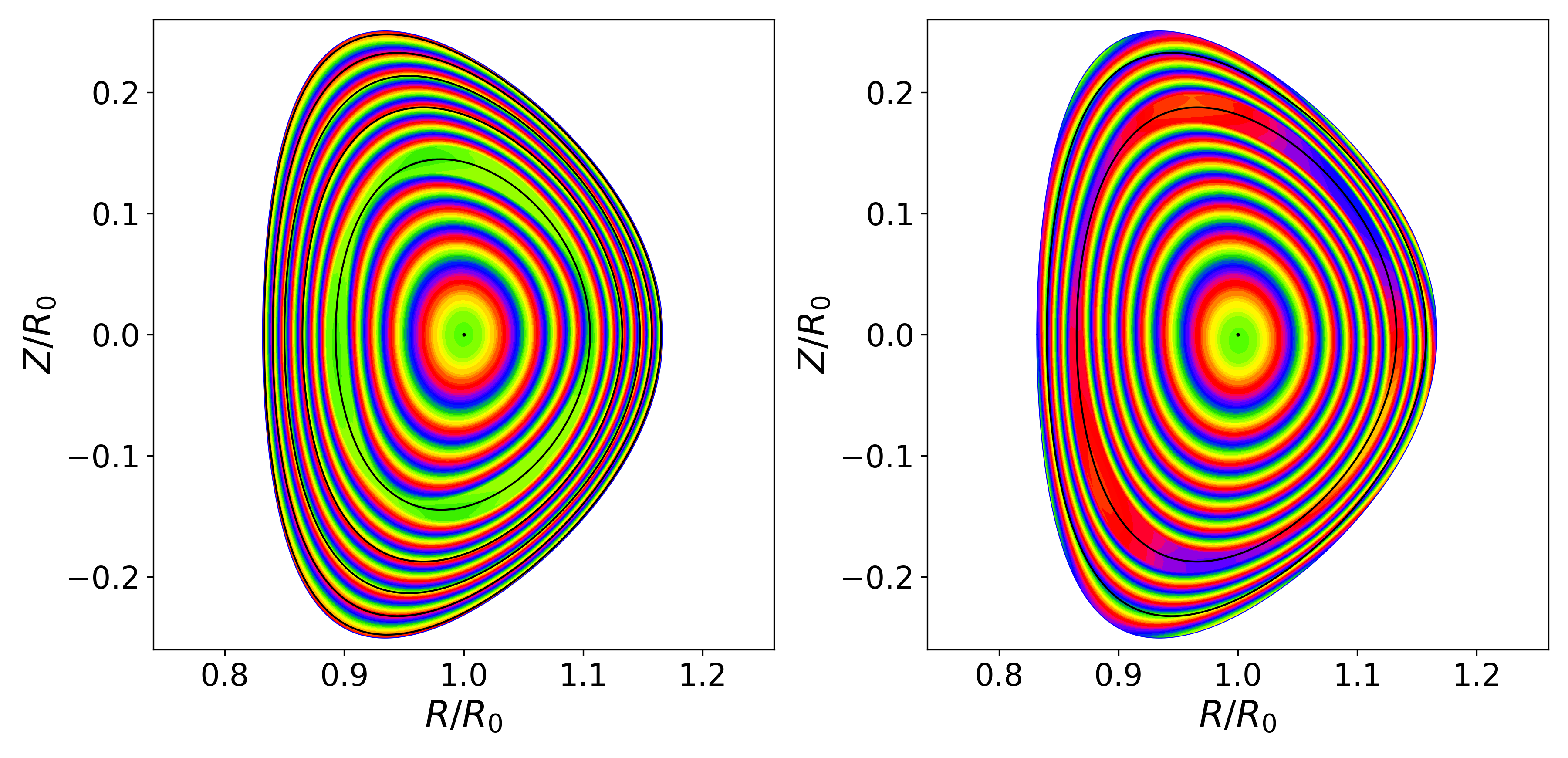}}
\caption{Total electron temperature  at a particular toroidal angle (which is the same as that in Fig.~\ref{fig11}) associated with a 3, 2 (left panel) and a 2, 1 (right panel) NTM of island width $W=0.1\,a$  in the example plasma equilibrium pictured in Figs.~\ref{figa}
and \ref{figb}. The black curves show the locations of the rational surfaces. The black dot shows the location of the magnetic axis.\label{fig12}}
\end{figure}

Figure~\ref{fig11} shows
the perturbed electron temperatures associated with both NTMs at a particular toroidal angle. In this case, all harmonics are included in the calculation. (The allowed poloidal harmonics in the
outer region have poloidal mode numbers in the range $m=-10$ to $m=+20$. The allowed overtone harmonics in the inner region are such that
$\nu$ lies in the range $0$ to 15.) It can be seen that the temperature perturbations have quite complicated structures. Nevertheless, when the temperature perturbations are
added to the equilibrium electron temperature profile then flat spots are clearly visible in the vicinity of the NTM rational surfaces, as shown in Fig.~\ref{fig12}. 

\section{Synthetic ECE Diagnostic}\label{s6}
\subsection{Introduction}
The ECE signal generated by an NTM is  measured on a horizontal chord that passes through the magnetic axis of the plasma. See Fig.~\ref{figa}.  Such a chord corresponds to a possible  path of ECE because 
the perpendicular gradient of the plasma refractive index is zero along the chord due to the up-down symmetric nature of the plasma equilibrium. The fact that the
chord path is normal to the equilibrium magnetic field means that Doppler
broadening of the ECE signal is eliminated.\cite{ece4a,ece5,bornatici} 

Gyrating electrons emit electron cyclotron radiation at frequencies that satisfy\,\cite{bornatici}
\begin{equation}
\omega= j\,\omega_c,
\end{equation}
where the positive integer $j$ is the harmonic number, and $\omega_c$ is the local cyclotron frequency. The cyclotron radiation propagates  along the
measurement chord in a direction perpendicular to the equilibrium magnetic field. There are two independent polarizations. O-mode radiation is
polarized such that the electric component of the wave is parallel to the equilibrium magnetic field, whereas X-mode radiation is polarized such that the
electric component of the wave is perpendicular to the field.\cite{plasma}

Neglecting the relativistic mass increase of electrons, the {\em cyclotron frequency}\/ takes the form\,\cite{plasma}
\begin{equation}
\omega_c = \frac{e\,B}{m_e},
\end{equation}
where $B$ is the local magnetic field-strength, $e$ the magnitude of the electron charge, and $m_e$ the electron rest mass.
Now, in ITER-like plasmas, it is an excellent approximation to write
\begin{equation}
B(R)= \frac{B_0\,R_0}{R}.
\end{equation}
Thus, it is convenient to define
\begin{equation}\label{romega}
R_\omega(\omega) = \frac{j\,\omega_{c\,0}\,R_0}{\omega},
\end{equation}
as the major radius from which $j$th harmonic ECE of frequency $\omega$ would originate in the absence of the relativistic mass increase. Here, $\omega_{c\,0}=e\,B_0/m_e$. 

\begin{figure}
\centerline{\includegraphics[width=0.7\textwidth]{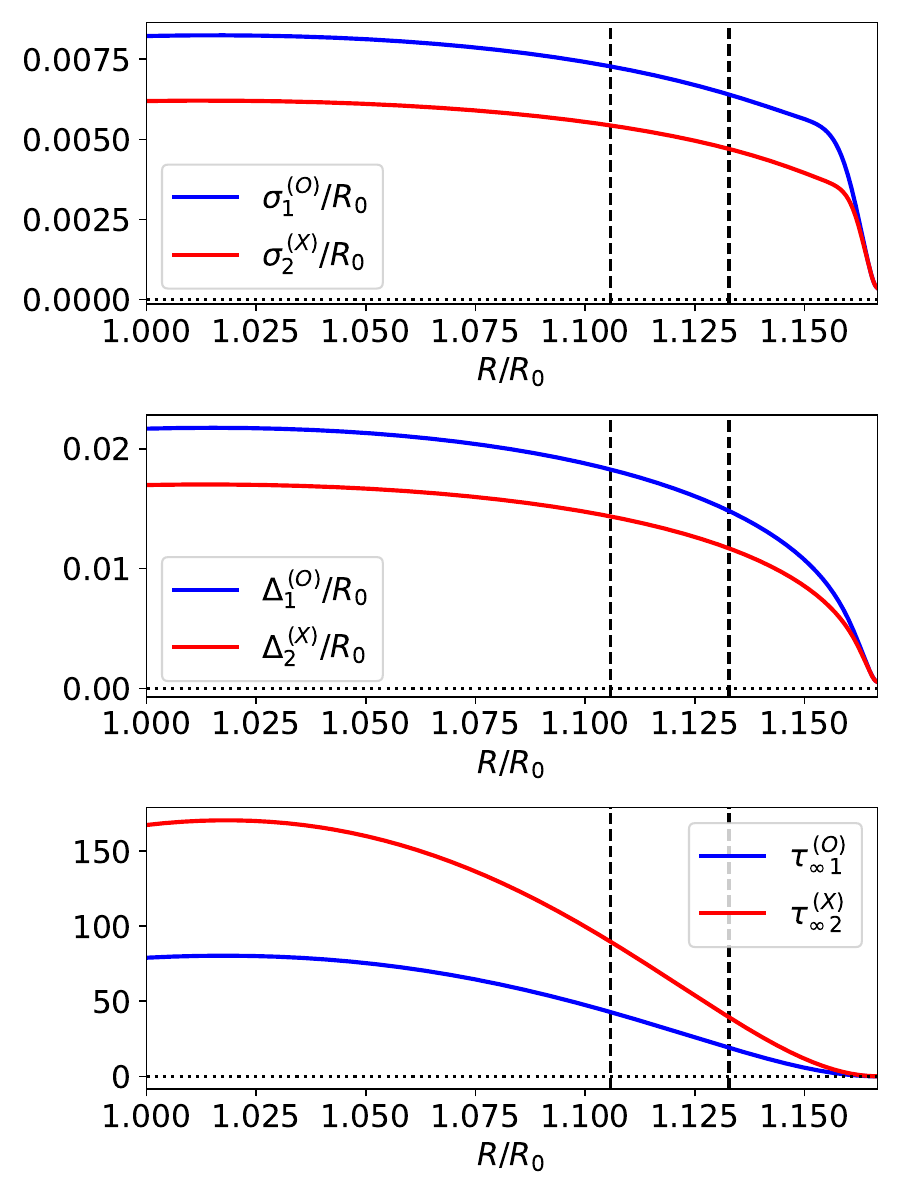}}
\caption{Standard deviation, $\sigma$, and inward radial shift, ${\mit\Delta}$, of the ECE spatial convolution function,  calculated as
a function of position on the measurement chord, for 1st harmonic O-mode and 2nd harmonic X-mode emission  in the model plasma equilibrium pictured in Figs.~\ref{figa} and \ref{figb}. Also, shown are the
saturated optical depths, $\tau_\infty$, of the two modes. The vertical dashed lines show the locations of the 3, 2 and the 2, 1 rational surfaces.}\label{fece}
\end{figure}

In thermal equilibrium, assuming that the electron number density is sufficiently large to render the plasma optically thick (see Fig.~\ref{fece}), the intensity of ECE  is directly proportional to the electron temperature, $T_e$,  of the region of the plasma from which the radiation originates.\cite{bornatici} 
 Moreover, the frequency of the emission, $\omega$, can be used to infer the major radius, $R_\omega(\omega)$,  of the
emitting region by means of Eq.~(\ref{romega}). In other words, an ECE diagnostic is capable of directly measuring the function $T_e(R_\omega)$ along the measurement chord.

\subsection{Relativistic Mass Increase}
 Unfortunately, the simple scheme just outlined is spoiled by the
fact that, due to the relativistic mass increase of gyrating electrons,  the true cyclotron frequency is
\begin{equation}
\omega_c = \frac{e\,B\,\sqrt{1-v^2/c^2}}{m_e},
\end{equation}
where $v$ is the electron speed, and $c$ the velocity of light in vacuum. Given that the gyrating electrons have a range of different speeds, they emit
cyclotron radiation in a range of frequencies that lies below the non-relativistic cyclotron frequency. This implies that radiation of a given frequency
actually emanates from a range of major radii that lie slightly inside $R_\omega$. The relativistic downshifting and broadening
in frequency of ECE  causes the temperature profile inferred from the diagnostic to be both shifted outward and smeared out in major radius. 

\subsection{Reabsorption} 
Now, gyrating electrons that emit electron cyclotron radiation of a given frequency are also capable of absorbing such radiation. 
 The reabsorption of the electron cyclotron radiation, as it propagates along the measurement chord, reduces the  downshifting and broadening
in frequency, and, therefore, increases the spatial resolution of the ECE diagnostic. The two ECE modes that are most strongly
absorbed by the plasma are 1st harmonic (i.e., $j=1$) O-mode, and 2nd harmonic (i.e., $j=2$) X-mode.\cite{bornatici}
Not surprisingly, the ITER ECE diagnostic is designed  to detect NTMs by means of these two modes.\cite{ece4a,ece5}

\subsection{Spatial Convolution Function}
The theory of the emission and reabsorption of electron cyclotron radiation is very complicated, and is summarized in Appendix~\ref{sece}. 
The net result of this theory is that the radiation temperature, $T_{\rm rad}(R_\omega)$, measured by the ECE diagnostic, is related
to the actual electron temperature along the measurement chord, $T_e(R)$, according to
\begin{equation}
T_{\rm rad}(R_\omega)  = \int^{R_{\rm lfs}}_{R_{\rm hfs}} T_e(R)\,F(R_\omega, R)\,dR.
\end{equation}
Here, $R_{\rm hfs}$ is the radius
of the plasma boundary on the high-field side (i.e., the inner side) of the chord, whereas $R_{\rm lfs}$ is the corresponding radius 
on the low-field side (i.e., the outer side). The {\em spatial convolution function}, $F(R_\omega,R)$, is well approximated as
\begin{align}
F(R_\omega,R)&= \frac{1}{\sqrt{2\pi}\,\sigma \,P({\mit\Delta}/\sigma)}\,
\exp\left[-\frac {(R-R_\omega+{\mit\Delta})^2}{2\,\sigma^2}\right]&R\leq R_\omega,\\[0.5ex]
&= 0& R>R_\omega.
\end{align}
Here, $P(x)$ is defined in Eq.~(\ref{Pxdef}). Note that the convolution function is characterized by by just two parameters: the standard deviation, $\sigma$, and the inward shift
of the function peak from the non-relativistic cyclotron resonance, ${\mit\Delta}$. 

\subsection{Results}
The analysis of Appendix~\ref{sece} allows us to calculate ${\mit\sigma}$ and ${\mit\Delta}$ for 1st harmonic O-mode and 2nd harmonic X-mode ECE along the
measurement chord of our example plasma equilibrium. The results of this calculation are shown in Fig.~\ref{fece}. Also shown are the saturated optical
depths, $\tau_\infty$, of the two modes. [See Eq.~(\ref{tinf}).] It can be seen that the plasma is optically thick (i.e., $\tau_\infty\gg 1$) to both
1st harmonic O-mode and second harmonic X-mode electron cyclotron radiation at the 3, 2 and the 2, 1 rational surfaces. However, the optical thickness of X-mode
radiation exceeds that of O-mode. Consequently, the standard deviation of the X-mode convolution function is less than that of the O-mode
function, indicating that the spatial resolution of an ECE diagnostic that employs the former mode is greater than that of one that employs the latter. 
It can be seen that ${\mit\Delta}\simeq 3\,\sigma$ for both modes. This indicates that the outward radial shift in the inferred location of
ECE emission, due to the relativistic mass increase, is a stronger effect than the radial smearing out of the emission. 

Figures \ref{fig16} and \ref{fig17} show the perturbed and total electron temperatures along the ECE measurement chord, as well as the corresponding temperatures inferred by a relativistically 
downshifted and broadened 1st harmonic O-mode and 2nd harmonic X-mode  ECE diagnostic, for a
3, 2 NTM and a 2, 1 NTM of island width $W=0.1\,a$,  in the model plasma equilibrium pictured in Figs.~\ref{figa} and \ref{figb}. 
The signals are evaluated at two different toroidal angles. However, because the
NTM temperature perturbation is actually rotating toroidally (because the NTM island chain is rotating), different toroidal angles translate to different measurement times. It can be seen that the temperature profile measured by the relativistically downshifted and 
broadened ECE signal is indeed both shifted outward and  smeared out in major radius. This effect is slightly more marked for O-mode compared to X-mode radiation. 

The flattening of the electron temperature in the vicinity of an NTM island chain is usually detected in tokamak experiments using the so-called ``Berrino algorithm", which looks for
a time-averaged anti-correlation between the ECE signal seen in neighboring channels.\cite{ece4}  We can  reproduce this algorithm  by calculating the radial gradient of
the relativistically downshifted and broadened ECE signals shown in Fig.~\ref{fig16}, via finite differencing,  and then averaging over toroidal angle. (In practice, we
calculate the signal at 32 equally-spaced toroidal angles and average the results.) Figure~\ref{fig17} shows our simulated Berrino algorithm. It can be seen that the
flattening of the electron temperature in the vicinity of an NTM island chain generates a local minimum in a signal that is otherwise monotonically increasing with major radius (except very close to
the edge of the plasma). It is clear from the figure that the algorithm is quite capable of detecting island chains of widths as small as 1\% of the plasma minor radius. 
However, if the location of the NTM rational surface is associated with the position of the local minimum of the signal, as is common practice,\cite{ece4}  then an error is introduced, because the
local minimum lies outside the rational surface due to the relativistic downshifting of the ECE. The fact that the algorithm can detect very narrow NTM island chains means that the stabilizing ECCD can be turned on before the chain causes significant
degradation of the plasma energy confinement. However, given that the ECCD ideally needs to hit one of the island O-points, which usually lie inside the associated rational 
surface, the radial location of the O-point inferred from the Berrino algorithm clearly needs to be corrected for both the relativistic downshifting and the fact that the O-point
is located slightly inside the rational surface. Figure~\ref{fig16} shows a corrected Berrino algorithm  in which the inferred major
radius of the signal is taken to be $R_\omega-{\mit\Delta}-\delta\,W/\sqrt{8}$, rather than $R_\omega$. In this case, the  local minima of the signals lie almost exactly
at the island O-points. 
 
\section{Summary and Discussion}\label{s7}
In this paper, we have demonstrated how a toroidal tearing mode code can be used to make realistic predictions of  the electron cyclotron emission (ECE) signals generated by  neoclassical
tearing modes (NTMs) in an ITER-like tokamak plasma equilibrium. (See Sect.~\ref{s2}.) In the outer region, which comprises the bulk of the plasma, helical
harmonics of the magnetic field with the same toroidal mode number as the NTM, but different poloidal mode numbers, are coupled together linearly by the Shafranov shift and
shaping of the equilibrium magnetic flux-surfaces. (See Sect.~\ref{s3}.) In the inner region, which is localized in the vicinity of the NTM rational surface, helical harmonics whose
poloidal and toroidal mode numbers are in the same ratio as those of the NTM are coupled together nonlinearly to produce a radially asymmetric magnetic
island chain. (See Sect.~\ref{s4}.)  The solutions in the inner and outer regions are asymptotically matched to one another. (See Sect.~\ref{s5}.) The asymptotic matching process determines  the
overall magnetic structure of the NTM, as well as the global perturbation to the electron temperature caused by the NTM. 

Although an NTM can easily be detected using magnetic
pickup coils located outside the plasma, this detection method is subject to substantial interference from other MHD modes, such
as sawtooth oscillations and ELMs, and also does not determine the location of the NTM island chain. This is problematic because the
cure for NTMs is to launch electron cyclotron  waves into the plasma in such a manner as  to drive a localized toroidal current in the vicinity of one of the O-points of the chain. 
In order to be effective, the location of the peak of the electron cyclotron current drive (ECCD) profile cannot miss the island O-point in major radius by more than two standard deviations of the radial width of the profile. (See Sect.~\ref{s5}.) Fortunately, as is confirmed  in this paper, ECE radiometry is capable of measuring the major radius of the island O-point, even when the width of the island chain is
very much less than the minor radius of the plasma. (See Sect.~\ref{s6}.) Moreover, this measurement is not
subject to interference from sawtooth oscillations or ELMs, because such modes do not produce flat spots in the
electron temperature profile.  However, the accuracy of the ECE diagnostic is limited by the fact that the signal is downshifted and
broadened in frequency due to the relativistic mass increase of the emitting electrons. It is clear, from the calculations in this paper, that the radial location of the
O-point deduced by an ECE diagnostic needs to be corrected for the relativistic downshifting, which causes the inferred location to shift outward in major radius from the rational surface, 
as well as the fact that the island O-point is shifted inward from the rational surface. An example of such a correction is shown in Fig.~\ref{fig16}. 

The calculations described in this paper were performed using the TJ toroidal tearing mode code,\cite{tear9,tear10} which is limited to unrealistically small aspect-ratio plasma equilibria. However, the
analysis in this paper could be incorporated into a toroidal tearing mode code, such as STRIDE, that can deal with realistic plasma equilibria, in a fairly
straightforward manner. Such an augmented code could calculate the necessary corrections to the Berrino algorithm very rapidly, and would constitute a
valuable addition to the ITER plasma control system. 

\section*{Acknowledgements}
This research was directly funded by the U.S.\ Department of Energy, Office of Science, Office of Fusion Energy Sciences, under  contract DE-SC0021156. 
The author acknowledges useful and informative discussions with W.L.~Rowan, J.P.~Ziegel, and Max Austin.

\section*{Data Availability Statement}
The digital data used in the figures in this paper can be obtained from the author upon reasonable request. The TJ code is freely 
available at {\tt https://github.com/rfitzp/TJ}. 

\appendix
\section{Electron Cyclotron Emission}\label{sece}
\subsection{Introduction}
This appendix reviews the theory of electron cyclotron emission (ECE), and is based on the analysis of Bornatici et al.,\cite{bornatici}
which is ultimately based on the work of Bekefi.\cite{bekefi}

\subsection{Orderings}
Our analysis premised on three main assumptions. 

Our first assumption is that the ECE emission lies in the so-called {\em relativistic regime}, in which the
broadening of the emission is predominately due to the relativistic mass increase of the emitting electrons, as opposed to the Doppler effect. The
emission is in the relativistic regime provided that
\begin{equation}\label{a1}
|N\,\cos\vartheta|  \ll \frac{v_t}{c},
\end{equation}
where
\begin{equation}
v_t = \left(\frac{T_e}{m_e}\right)^{1/2}.
\end{equation}
Here, $N$ is the refractive index of the plasma, $\vartheta$ the angle subtended between the ECE wave-vector and the equilibrium magnetic field, $T_e$ the equilibrium electron
temperature, $m_e$ the electron rest mass, $c$ the velocity of light in vacuum, and $v_t$ the electron thermal speed.  All quantities are evaluated at the ECE emission site, which invariably
lies 
close to the point at which the cyclotron resonance condition
\begin{equation}
\omega = j\,\omega_c
\end{equation}
is satisfied, 
where 
\begin{equation}
\omega_c = \frac{e\,B}{m_e}.
\end{equation}
Here, $\omega$ is the angular frequency of the emitted radiation, $e$ the magnitude of the electron charge, $B$ the equilibrium magnetic field-strength at the emission site, and $\omega_c$  the local (non-relativistic) cyclotron frequency. 
Moreover, $j=1$ for 1st harmonic emission, whereas $j=2$ for 2nd harmonic emission. In fact, the constraint (\ref{a1}) is easily satisfied because we are assuming that
$\vartheta=\pi/2$. 

Our second assumption is that the ECE emission lies in the so-called {\em small gyro-radius regime}, in which the gyro-radii of the emitting electrons are much smaller than
the wavelength of the emitted radiation. This regime holds provided that 
\begin{equation}
j^{\,2}\,N^2\,\left(\frac{v_t}{c}\right)^2 = 1.96\times 10^{-3}\,j^{\,2}\,N^2\,T_e({\rm keV}) \ll 1,
\end{equation}
which is easily satisfied in ITER-like plasmas. 

Our final assumption is that the ECE emission lies in the so-called {\em tenuous plasma regime}, in which
\begin{equation}\label{a6}
\left(\frac{\omega_p}{\omega_c}\right)^2 = 10.3\,\frac{n_e(10^{20}\,{\rm m}^{-3})}{B^2({\rm T})}\ll 1,
\end{equation}
where
\begin{equation}
\omega_p = \left(\frac{n_e\,e^2}{\epsilon_0\,m_e}\right)^{1/2}
\end{equation}
is the electron plasma frequency at the emission site, whereas $n_e$ is the local electron number density. The constraint (\ref{a6}) is
easily satisfied in the ITER-like tokamak plasma equilibrium considered in this paper. 

\subsection{ECE Signal}
The ECE signal generated by an NTM is  measured on a horizontal chord that passes through the magnetic axis of the plasma. See Fig.~\ref{figa}. 
In thermal equilibrium, assuming that the plasma is optically thick to ECE radiation (see Figs.~\ref{Omode} and \ref{Xmode}), the so-called {\em radiation temperature}\/ measured by the ECE diagnostic is related to the {\em radiance}, $I(\omega)$, of the ECE signal as follows:\,\cite{bornatici}
\begin{equation}\label{a8}
T_{\rm rad}(\omega)=\frac{8\pi^3\,c^{\,2}\, I(\omega)}{\omega^{\,2}}.
\end{equation}
Here,
\begin{equation}\label{a9}
I(\omega) = \frac{\omega^{\,2}}{8\pi^3\,c^{\,2}}\int_{R_{\rm hfs}}^{R_{\rm lfs}}G(\omega, R)\,dR,
\end{equation}
where $G(\omega,R)$ is known as the {\em emissivity function}. Moreover, $R$ is the major radius of a location on the ECE measurement chord, $R_{\rm hfs}$ is the radius
of the plasma boundary on the high-field side (i.e., the inner side) of the chord, whereas $R_{\rm lfs}$ is the corresponding radius 
on the low-field side (i.e., the outer side).

The emissivity function takes the form\,\cite{bornatici}
\begin{equation}\label{a10}
G(\omega, R) = T_e(R)\,\alpha(\omega,R)\,\exp\left[-\tau(\omega,R)\right],
\end{equation}
where 
\begin{equation}\label{a11}
\tau(\omega,R) = \int_R^{R_{\rm lfs}}\alpha(\omega,R')\,dR'
\end{equation}
is the dimensionless {\em optical depth}. 
Here, $T_e(R)$ is the electron temperature on the measurement chord (including the perturbation due to the NTM), whereas $\alpha(\omega,R)$
is termed the {\em absorption coefficient}.

In ITER-like plasmas, it is an excellent approximation to write
\begin{equation}
B(R) = \frac{B_0\,R_0}{R},
\end{equation}
where $B_0$ is the toroidal magnetic field-strength at the magnetic axis.  It follows that
\begin{equation}\label{a13}
\omega_c(R) = \frac{\omega_{c\,0}\,R_0}{R},
\end{equation}
where
\begin{equation}
\omega_{c\,0} = \frac{e\,B_0}{m_e}
\end{equation}
is the (non-relativistic) cyclotron frequency on the magnetic axis. Thus, we can use the local (non-relativistic) cyclotron frequency, $\omega_c(R)$, as a proxy for major radius, $R$,  along the measurement chord. 

Equations~(\ref{a8})--(\ref{a11}) and (\ref{a13}) can be combined to give
\begin{equation}\label{a15}
T_{\rm rad}(\omega) = \int_{\omega_{c\,{\rm lfs}}}^{\omega_{c\,{\rm hfs}}}T_e(\omega_c)\,H(\omega,\omega_c)\,d\omega_c,
\end{equation}
where
\begin{align}
H(\omega,\omega_c)&=\tau_0\,\frac{\hat{\alpha}(\omega,\omega_c)}{\omega_c}\,\exp\left[-\tau(\omega,\omega_c)\right],\\[0.5ex]
\tau(\omega,\omega_c)&= \tau_0\int_{\omega/j}^{\omega_c}\frac{\hat{\alpha}(\omega,\omega_c')}{\omega_c'}\,d\omega_c',
\end{align}
Here, $\omega_{c\,{\rm lfs}}= \omega_{c\,0}\,R_0/R_{\rm lfs}$,  $\omega_{c\,{\rm hfs}}= \omega_{c\,0}\,R_0/R_{\rm hfs}$, 
\begin{equation}
\tau_0 = \frac{\omega_{c\,0}\,R_0}{c}= 5.97\times 10^2\,R_0({\rm m})\,B_0({\rm T})
\end{equation}
is the dimensionless  ratio of the major radius of the plasma to the typical wavelength of the ECE, and $\hat{\alpha}(\omega,\omega_c) = (c/\omega_c)\,\alpha(\omega,\omega_c)$
is a convenient dimensionless form of the absorption coefficient. 

 It is clear from Eq.~(\ref{a15}) that the radiation temperature measured by the ECE diagnostic, $T_{\rm rad}(\omega)$,  is a convolution of the true
electron temperature, $T_e(\omega)$,  with the {\em spectral convolution function}, $H(\omega,\omega_c)$. The convolution function specifies the downshifting and broadening   (in frequency) of the ECE
signal due to the relativistic mass increase of the emitting electrons.

It is convenient to define
\begin{equation}
R_\omega(\omega) = \frac{j\,\omega_{c\,0}\,R_0}{\omega},
\end{equation}
as the major radius from which $j$th harmonic ECE would originate in the absence of the aforementioned downshifting and broadening of the signal. 
Thus, we can also write
\begin{equation}\label{a20}
T_{\rm rad}(R_\omega) = \int^{R_{\rm lfs}}_{R_{\rm hfs}}T_e(R)\,F(R_\omega,R)\,dR,
\end{equation}
where
\begin{align}\label{a21}
F(R_\omega,R) &= \frac{1}{1-\exp[-\tau_\infty(R_\omega)]}\,\tau_0\,\frac{\hat{\alpha}(R_\omega,R)}{R}\,\exp[-\tau(R_\omega, R)],\\[0.5ex]
\tau(R_\omega,R) &= \tau_0\int_R^{R_\omega}\frac{\hat{\alpha}(R_\omega,R')}{R'}\,dR,\\[0.5ex]
\tau_\infty(R_\omega) &= \tau_0\int^{R_\omega}_{R_{\rm hfs}}\frac{\hat{\alpha}(R_\omega,R')}{R'}\,dR.\label{tinf}
\end{align}
According to Eq.~(\ref{a20}), the electron temperature profile along the measurement chord that is inferred from the ECE diagnostic, $T_{\rm rad}(R_\omega)$,
is the convolution of the true temperature profile, $T_e(R)$, with the {\em spatial convolution function}, $F(R_\omega,R)$. The first term on the right-hand side of
Eq.~(\ref{a21}) is a correction made to ensure that the area under the convolution function is unity. 

\subsection{First Harmonic O-Mode Absorption Coefficient} 
The dimensionless absorption coefficient for 1st harmonic O-mode ECE is\,\cite{bornatici}
\begin{equation}\label{a22}
\hat{\alpha}_1^{\,(O)}=\frac{N'\,(1/2)\,(\omega_p/\omega_c)^2\,[-F''_{7/2}(z_1)]}{1+(1/2)\,(\omega_p/\omega_c)^2\,F_{7/2}'(z_1)},
\end{equation}
where
\begin{equation}
N'^{\,2}=\frac{1-(\omega_p/\omega_c)^2}{1+(1/2)\,(\omega_p/\omega_c)^2\,F'_{7/2}(z_1)},
\end{equation}
and
\begin{equation}\label{a24}
z_1=\left(\frac{c}{v_t}\right)^2\left(1-\frac{\omega_c}{\omega}\right).
\end{equation}
Here, $N'$ is the real part of the plasma refractive index, and $\omega_p$ and $v_t$ are evaluated at $z_1=0$. Moreover,
$F_{7/2}'(z)$ and $F_{7/2}''(z)$ are, respectively,  the real and imaginary parts of
the function\,\cite{bornatici}
\begin{equation}
F_{7/2}(z) =\frac{8}{15}\left[z^2-\frac{z}{2}+\frac{3}{4}+{\rm i}\,z^{5/2}\,Z({\rm i}\sqrt{z})\right],
\end{equation}
where $Z(\xi)$ is the {\em plasma dispersion function}.\cite{plasma} The complex plasma dispersion function is most conveniently evaluated numerically 
in terms of the {\em Faddeeva function},\cite{faddeeva}
\begin{equation}
w(\xi) = \exp(-\xi^2)\,{\rm erfc}(-{\rm i}\,\xi).
\end{equation}
where ${\rm erfc}(z)$ is the complimentary error function,\cite{as} 
as follows:
\begin{equation}
Z(\xi)= {\rm i}\,\pi^{1/2}\,w(\xi).
\end{equation}

 Figure~\ref{f72} shows the real and imaginary parts of the function $F_{7/2}(z)$. It is clear from Eq.~(\ref{a22}),
(\ref{a24}), and the figure, that 1st harmonic O-mode  ECE is absorbed by the plasma in a range of frequencies that lie slightly above the local (non-relativistic) cyclotron frequency,
$\omega_c$. This corresponds to a range of major radii that lies just inside the location of the 1st harmonic (non-relativistic) cyclotron resonance. 

Figure~\ref{Omode} shows the normalized absorption coefficient, optical depth, spectral convolution function, and spatial convolution function for
1st harmonic O-mode ECE in a typical ITER-like plasma. It can be seen that the saturated optical depth is quite large (about 50), and that the full 
width of the spatial convolution function is about 10 cm. The dotted curve in the bottom right panel of the figure shows a fit to the
spatial convolution function. The fitting function is a modified Gaussian:
\begin{align}
F_1^{\,(O)}(R_\omega,R)&= \frac{1}{\sqrt{2\pi}\,\sigma_1^{\,(O)} \,P({\mit\Delta}_1^{\,O}/\sigma_1^{\,(O)})}\,
\exp\left(-\frac{\left[R-R_\omega+{\mit\Delta}_1^{\,(O)}\right]^2}{2\left[\sigma_1^{\,(O)}\right]^2}\right)&R\leq R_\omega,\\[0.5ex]
&= 0& R>R_\omega,
\end{align}
where 
\begin{equation}\label{Pxdef}
P(x) = \frac{1}{\sqrt{2\pi}}\int_{-\infty}^x{\rm e}^{-y^2}\,dy.
\end{equation}
Note that the fitting function ensures that $F_1^{\,(O)}(R_\omega,R)= 0$ for $R>R_\omega$, in accordance with Eq.~(\ref{a21}). Moreover, the area under the
fitting function is unity. 
The fit is excellent, indicating that the convolution function 
 is characterized by just two parameters: the standard deviation, $\sigma_1^{\,(O)}$, and the inward shift
of the function peak from the non-relativistic cyclotron resonance, ${\mit\Delta}_1^{\,(O)}$. 

\subsection{Second Harmonic X-Mode  Absorption Coefficient}
The dimensionless absorption coefficient for 2nd harmonic X-mode ECE is\,\cite{bornatici}
\begin{equation}\label{a28}
\hat{\alpha}_2^{\,(X)} = \frac{N'\,(\omega_p/\omega_c)^2\,(1+a_2)^2\,[-F_{7/2}''(z_2)]}
{1+(1/2)\,(\omega_p/\omega_c)^2\,(1+a_2)^2\,F_{7/2}'(z_2)},
\end{equation}
where
\begin{align}
a_2 &= \frac{(1/6)\,(\omega_p/\omega_c)^2\left[1+3\,N'^{\,2}\,F_{7/2}'(z_2)\right]}{1- (1/3)\,(\omega_p/
\omega_c)^2\left[1+(3/2)\,N'^{\,2}\,F_{7/2}'(z_2)\right]},\\[0.5ex]
N'^{\,2} &= N_c^{\,2}\left[1-(b+a\,N_c^{\,2})\right],\\[0.5ex]
a&= - \frac{(1/2)\,(\omega_p/\omega_c)^2\,F'_{7/2}(z_2)}{1-(1/3)\,(\omega_p/\omega_c)^2},\\[0.5ex]
b &= -2\left[1-\frac{1}{6}\left(\frac{\omega_p}{\omega_c}\right)^2\right]a,\\[0.5ex]
N_c^{\,2}&= 1 -\frac{(1/3)\,(\omega_p/\omega_c)^2\,[1-(1/4)\,(\omega_p/\omega_c)^2]}{1-(1/3)\,(\omega_p/\omega_c)^2},\\[0.5ex]
z_2 &= \left(\frac{c}{v_t}\right)^2\left(1-\frac{2\,\omega_c}{\omega}\right).\label{a29}
\end{align}
Here, $v_t$ and $\omega_p$ are evaluated at $z_2=0$. Moreover, the parameters $a$, $b$, and $N_c$ have been calculated assuming that $\omega=2\,\omega_c$. 

 It is clear from Eq.~(\ref{a28}),
(\ref{a29}), and Fig.~\ref{f72}, that 2nd harmonic X-mode  ECE is absorbed by the plasma in a range of frequencies that lie slightly above twice the local (non-relativistic) cyclotron frequency. This corresponds to a range of major radii that lies just inside the location of the 2nd harmonic (non-relativistic) cyclotron resonance. 

Figure~\ref{Xmode} shows the normalized absorption coefficient, optical depth, spectral convolution function, and spatial convolution function for
2nd harmonic O-mode ECE in a typical ITER-like plasma. It can be seen that the saturated optical depth is very  large (about 100), and that the full 
width of the spatial convolution function is about 7 cm.

  The dotted curve in the bottom right panel of the figure shows a fit to the
spatial convolution function. As before, the fitting function is a modified Gaussian:
\begin{align}
F_2^{\,(X)}(R_\omega,R)&= \frac{1}{\sqrt{2\pi}\,\sigma_2^{\,(X)} \,P({\mit\Delta}_2^{\,X}/\sigma_2^{\,(X)})}\,
\exp\left(-\frac{\left[R-R_\omega+{\mit\Delta}_2^{\,(X)}\right]^2}{2\left[\sigma_2^{\,(X)}\right]^2}\right)&R\leq R_\omega,\\[0.5ex]
&= 0& R>R_\omega.
\end{align}
  The fit is very good, indicating that the convolution function 
is characterized by just two parameters: the standard deviation, $\sigma_2^{\,(X)}$, and the inward shift
of the function peak from the non-relativistic cyclotron resonance, ${\mit\Delta}_2^{\,(X)}$.

\newpage
\begin{table}
\begin{tabular}{ccccccccc}
\hline
$m$ & $n$ & $\hat{r}_l$ & $L_s$ & $E_{ll}$ & ${\mit\Delta}_{l\,{\rm crit}}$ & $\hat{\beta}$ & $\alpha_b$ & $\alpha_c$\\[0.5ex]\hline
~3~&~2~ & ~$0.6195$~ & ~$1.756$~ & ~$1.709$~ & ~$17.56$~ & ~$0.0540$~ & ~$2.588$~ & ~$0.5816$~\\[0.5ex]
2& 1& $0.7874$  & ~$1.209$~&$ 13.66$ & $47.96$ & $0.0293$ & $3.061$ & $1.000$\\\hline
\end{tabular}
\caption{The poloidal mode number, toroidal mode number, normalized minor radius, normalized magnetic shear-length, linear tearing stability index, critical linear tearing stability index, normalized
electron pressure, bootstrap parameter, and curvature parameter, respectively, at the $3$, $2$ and the $2$, $1$ rational surfaces of the example
plasma equilibrium pictured in Figs.~\ref{figa} and \ref{figb}. \label{t1} }
\end{table}

\begin{table}
\begin{tabular}{ccccccc}
\hline
$m$ & $n$ & $W/a$ & ${\mit\Psi}$ & $\delta$ & ~$\delta T_{e\,-}({\rm eV})$~ & $\delta T_{e\,+}({\rm eV})$ \\[0.5ex]\hline
3& 2& $0.1$  & ~$3.48\times 10^{-4}$~&$0.150$ &  $-1.90\times 10^{3}$ & $-0.276$ \\[0.5ex]
~1~&~2~ & ~$0.1$~ & ~$4.89\times 10^{-4}$~ & ~$0.272$~ &~$-2.39\times 10^{3}$~& ~$-3.91$ ~\\\hline
\end{tabular}
\caption{The poloidal mode number, toroidal mode number, island width, normalized reconnected flux, island asymmetry parameter, equilibrium electron temperature reduction
inside the rational surface,  and equilibrium electron temperature reduction
outside the rational surface, respectively, for a $3$, $2$ and a $2$, $1$ NTM in the example
plasma equilibrium pictured in Figs.~\ref{figa} and \ref{figb}.  \label{t2} }
\end{table}

\begin{figure}
\centerline{\includegraphics[width=\textwidth]{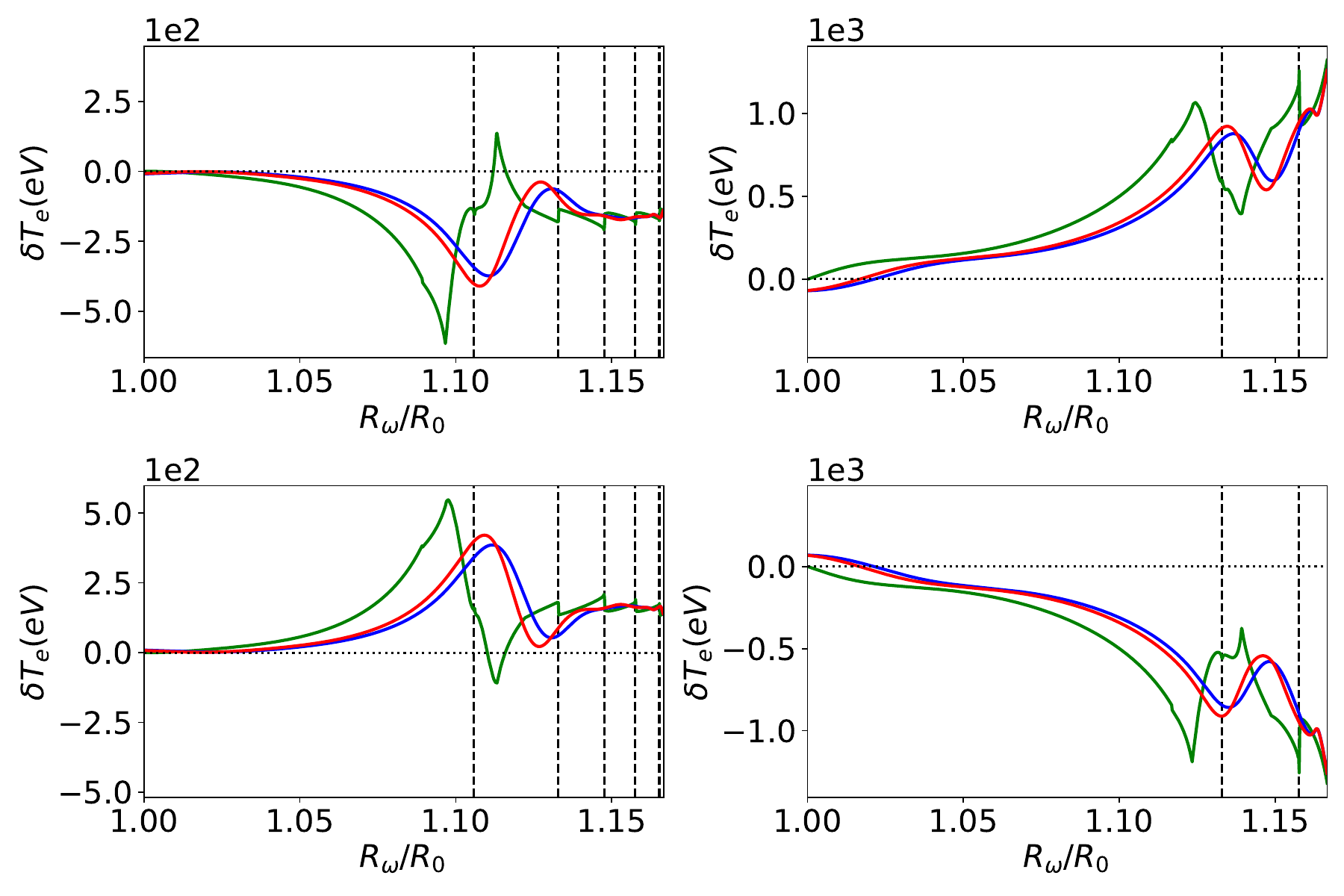}}
\caption{Perturbed electron temperature along the ECE measurement chord (green curves) and the perturbed temperature inferred by the relativistically downshifted and  broadened 1st harmonic
O-mode (blue curves) and 2nd harmonic X-mode (red curves) ECE diagnostic for a
3, 2 NTM (left panels) and a 2, 1 NTM (right panels) of island width $W=0.1\,a$ in the model plasma equilibrium pictured in Figs.~\ref{figa} and \ref{figb}.
The perturbed temperatures are calculated  at two different toroidal angles (top and bottom panels). 
The vertical dashed lines show the locations of the rational surfaces. 
\label{fig16}}
\end{figure}

\begin{figure}
\centerline{\includegraphics[width=\textwidth]{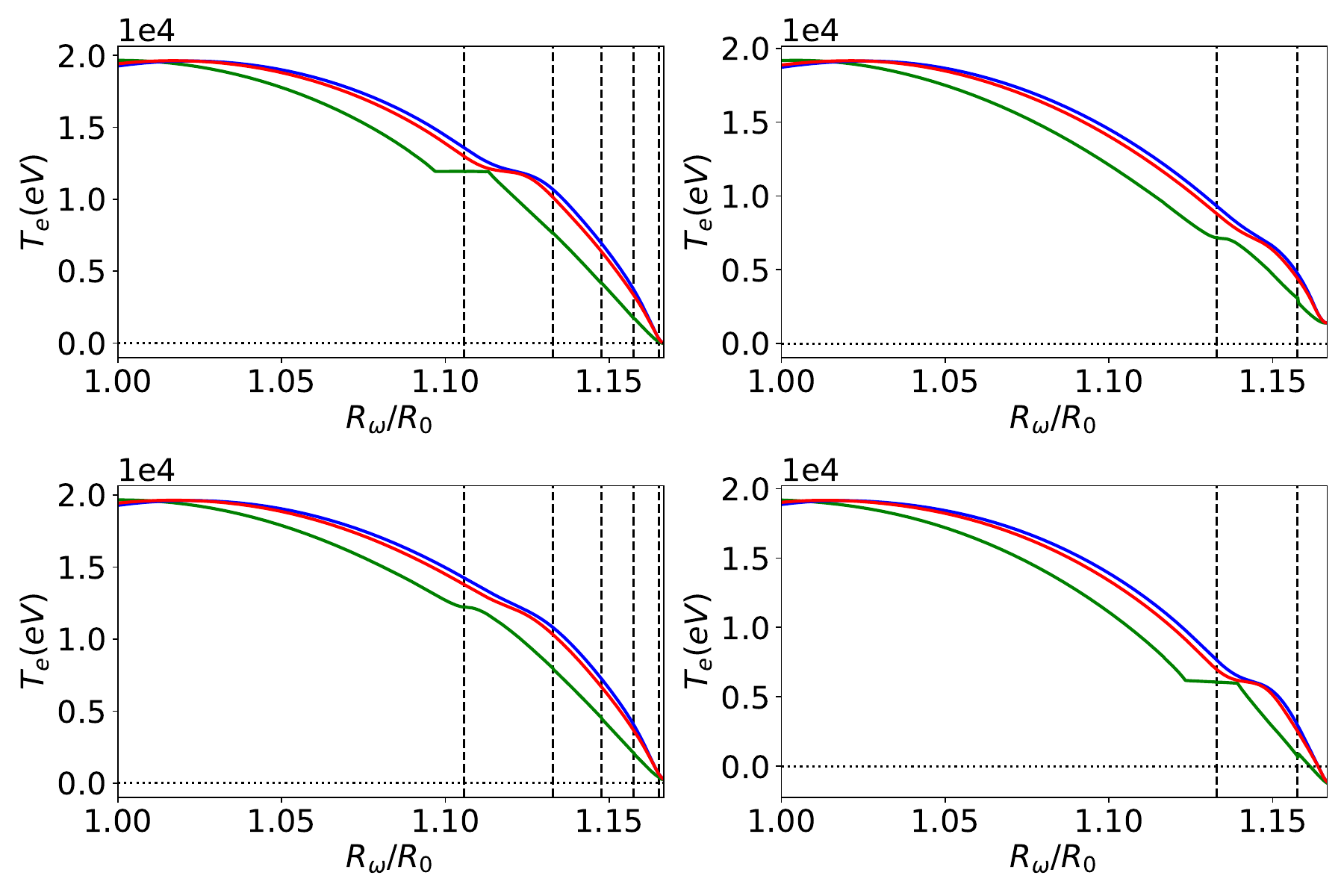}}
\caption{Total electron temperature along the ECE chord (green curves) and the perturbed temperature inferred by the relativistically downshifted and broadened 
 1st harmonic
O-mode (blue curves) and 2nd harmonic X-mode (red curves) ECE diagnostic  for a
3, 2 NTM (left panels) and a 2, 1 NTM (right panels) of island width $W=0.1\,a$  in the model plasma equilibrium pictured in Figs.~\ref{figa} and \ref{figb}.
The total temperatures are calculated  at two different toroidal angles (top and bottom panels).  The vertical dashed lines show the locations of the rational surfaces. \label{fig17}}
\end{figure}

\begin{figure}
\centerline{\includegraphics[width=\textwidth]{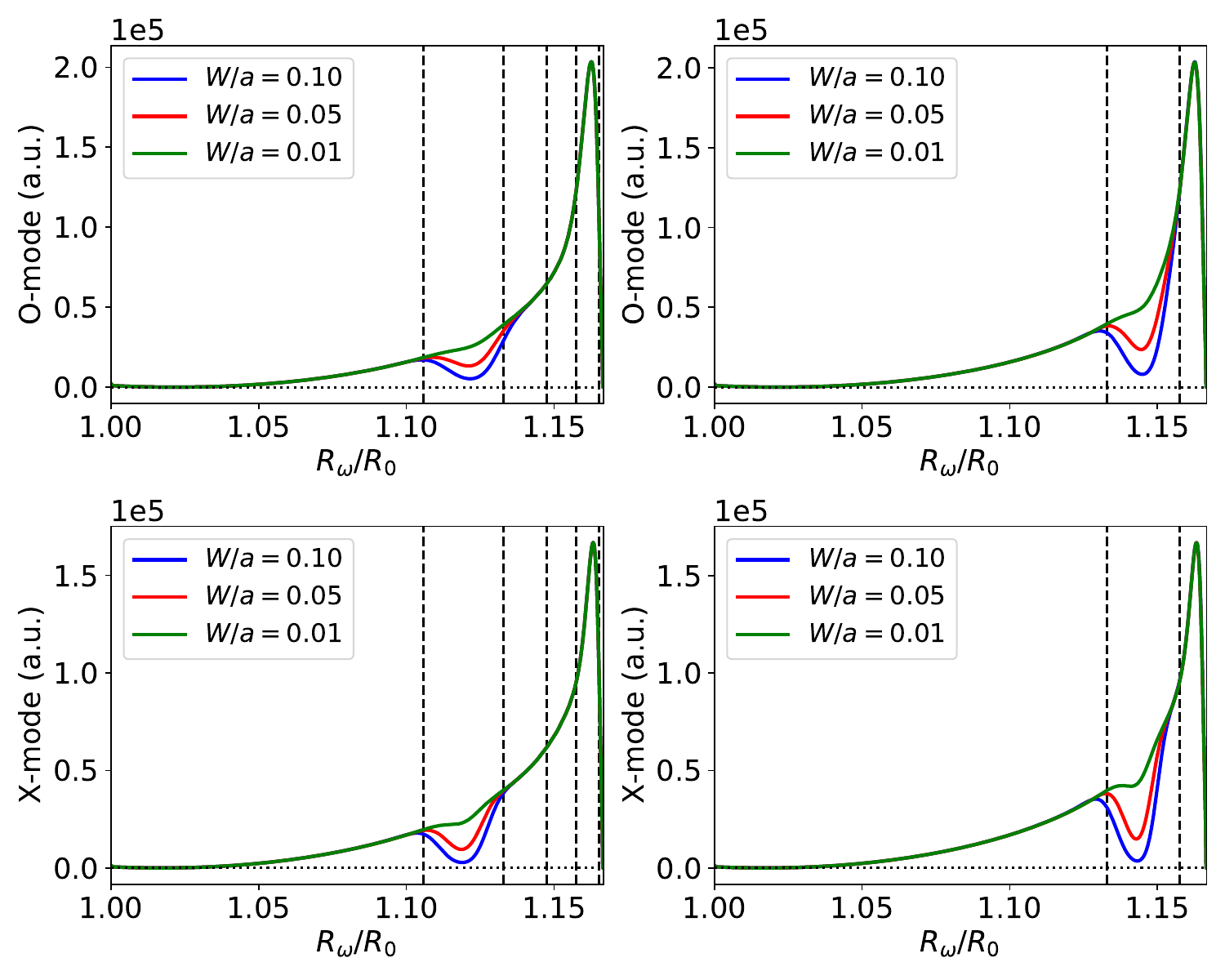}}
\caption{Implementation of the Berrino algorithm for detecting electron temperature flattening  in the vicinity of an NTM island chain  for a 3, 2  NTM (left panels) and a 2, 1 NTM (right panels) in the model plasma equilibrium pictured in Figs.~\ref{figa} and \ref{figb}. The top panels show the 1st harmonic O-mode signals,
whereas the bottom panels show the 2nd harmonic O-mode signals. The vertical dashed lines show the locations of the rational surfaces. \label{fig15}}
\end{figure}

\begin{figure}
\centerline{\includegraphics[width=\textwidth]{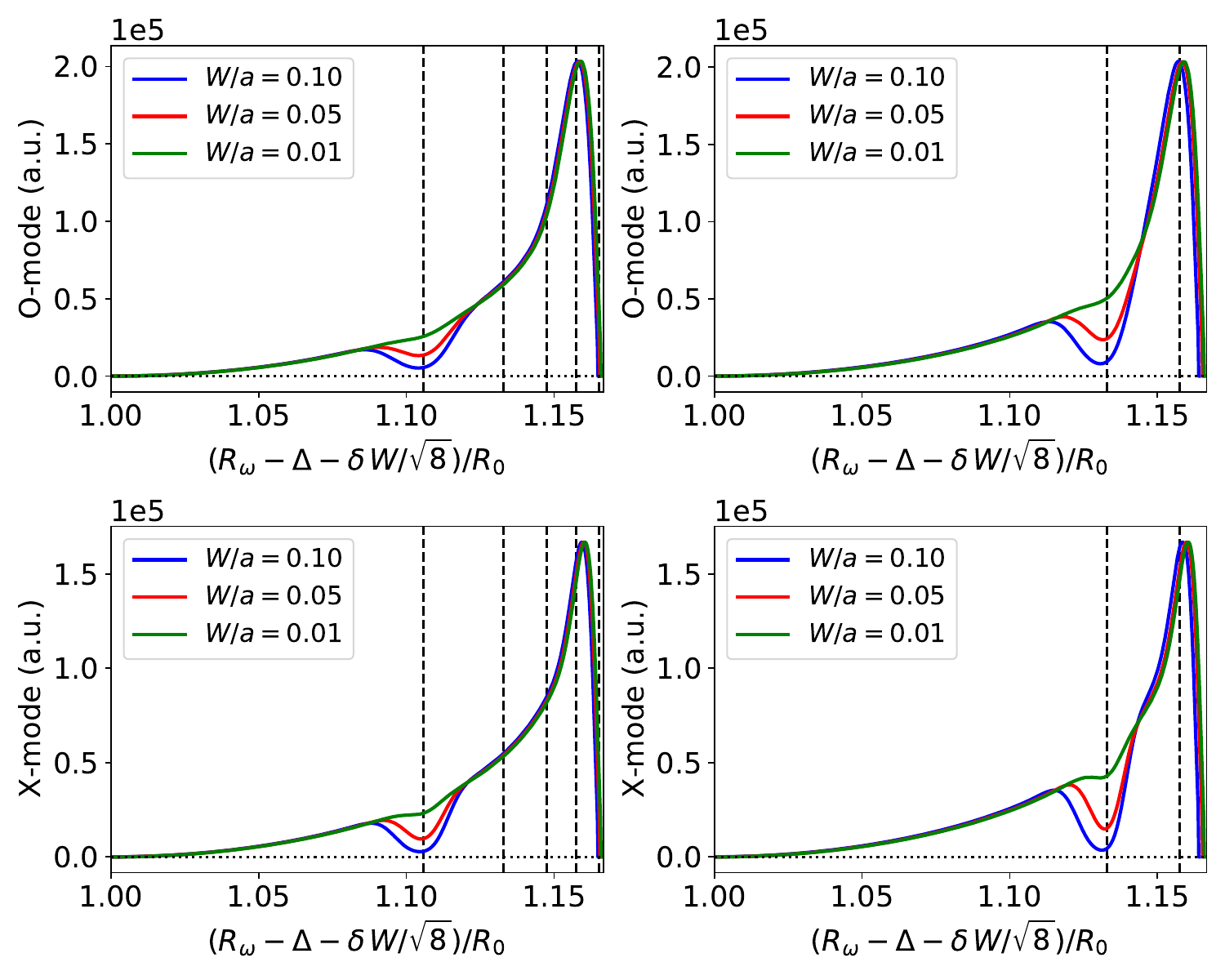}}
\caption{Implementation of a corrected  Berrino algorithm for detecting electron temperature flattening  in the vicinity of an NTM island chain  for a 3, 2  NTM (left panels) and a 2, 1 NTM (right panels) in the model plasma equilibrium pictured in Figs.~\ref{figa} and \ref{figb}. The top panels show the 1st harmonic O-mode signals,
whereas the bottom panels show the 2nd harmonic O-mode signals.The vertical dashed lines show the locations of the rational surfaces. \label{fig16}}
\end{figure}

\begin{figure}
\centerline{\includegraphics[width=\textwidth]{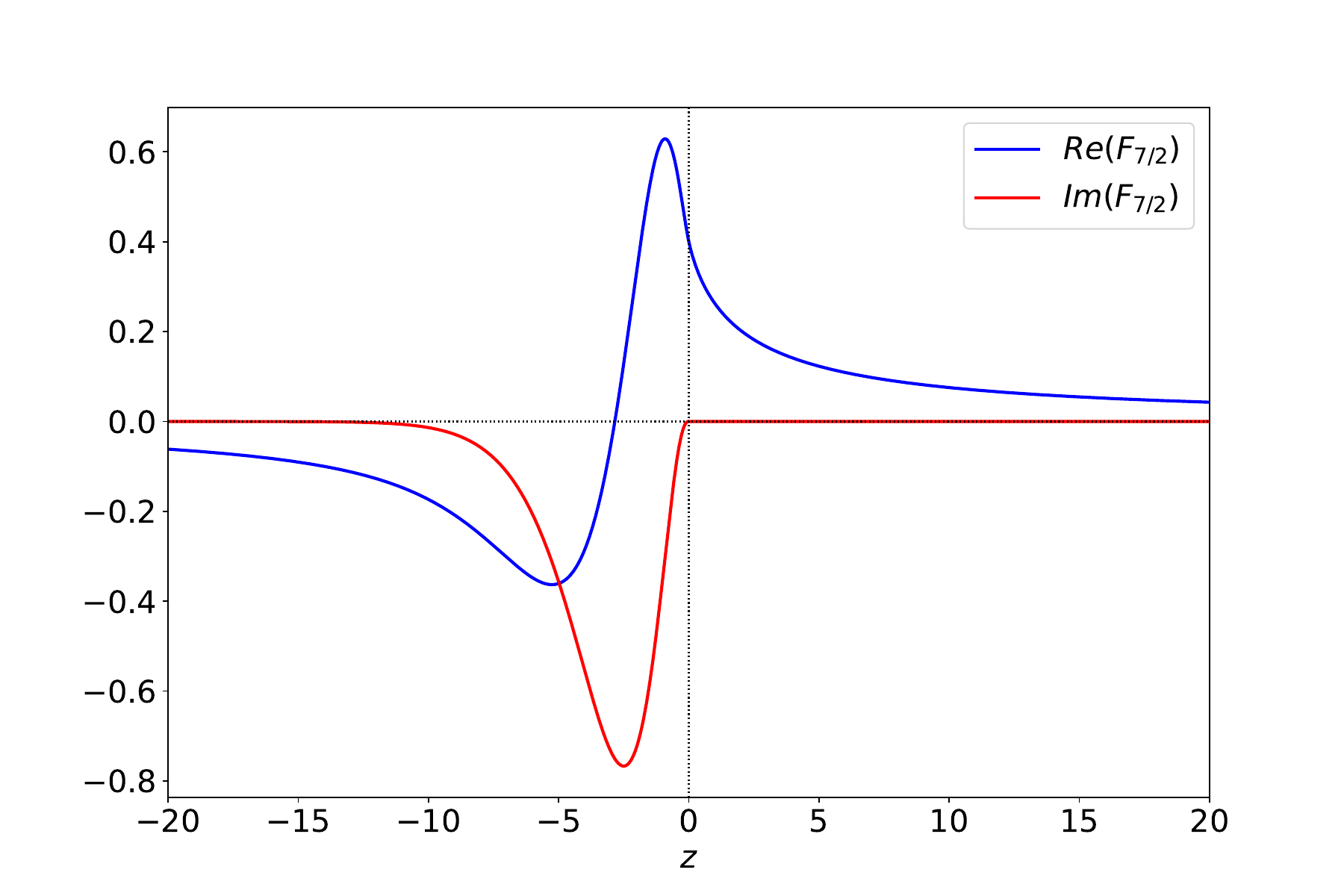}}
\caption{The real and imaginary parts of the function $F_{7/2}(z)$.\label{f72}}
\end{figure}

\begin{figure}
\centerline{\includegraphics[width=\textwidth]{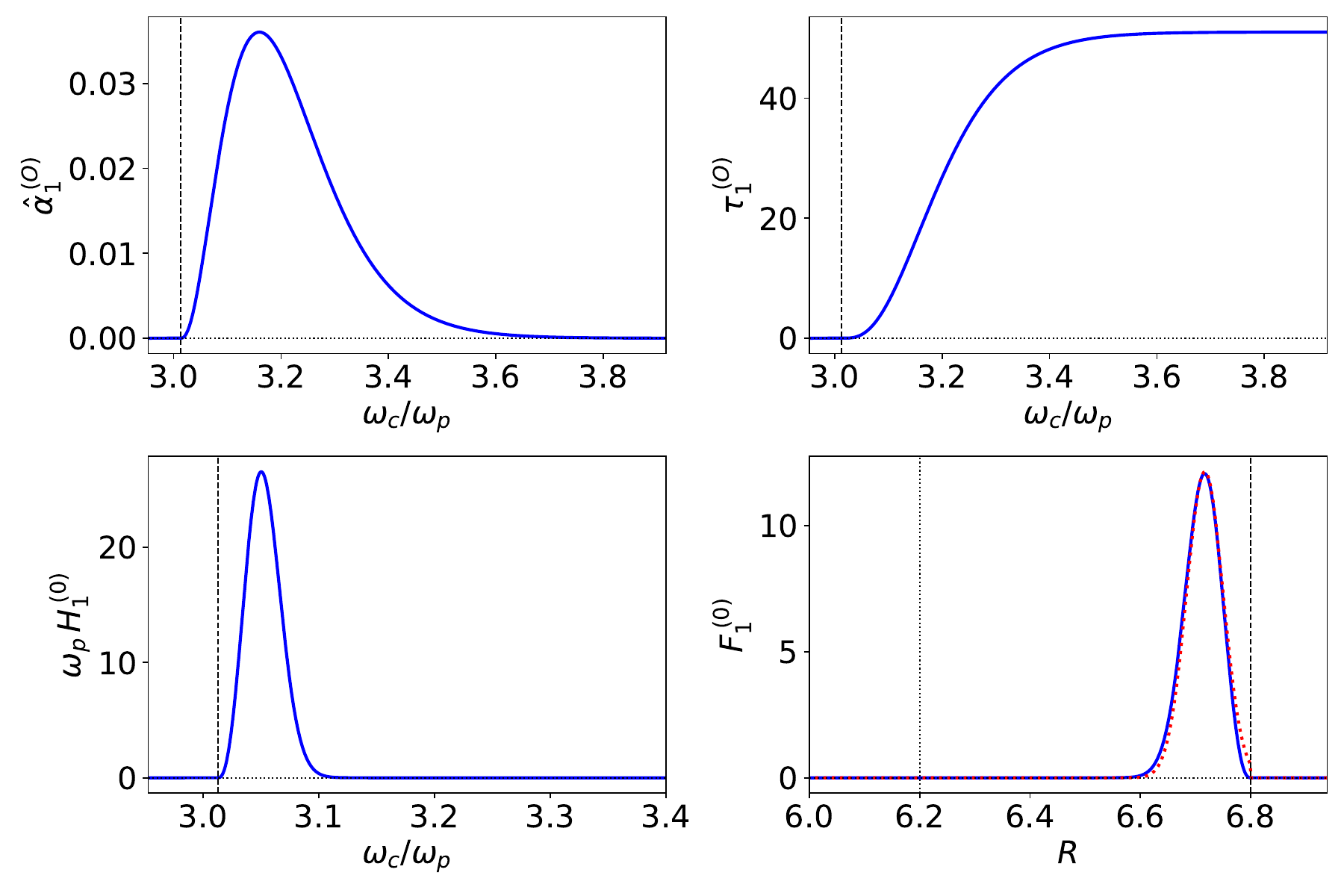}}
\caption{The normalized  absorption coefficient (top left), optical depth (top right), normalized
spectral convolution function (bottom left), and spatial convolution function (bottom right), for 1st harmonic O-mode ECE from an
ITER-like plasma characterized by  $T_e=10\,{\rm keV}$, $n_e=2.5\times 10^{19}\,{\rm m}^{-3}$ (at the 1st harmonic cyclotron resonance), $B_0=5.3\,{\rm T}$, $R_0=6.2\,{\rm m}$, and $R_\omega= 6.8\,{\rm m}$.
The vertical dashed lines show the location of the 1st harmonic cyclotron resonance.  The vertical dotted line (in the bottom right panel) shows the location of the  magnetic axis. 
The dotted red curve in the bottom right panel is a  fit to the true spatial convolution function (shown in blue).\label{Omode}}
\end{figure}

\begin{figure}
\centerline{\includegraphics[width=\textwidth]{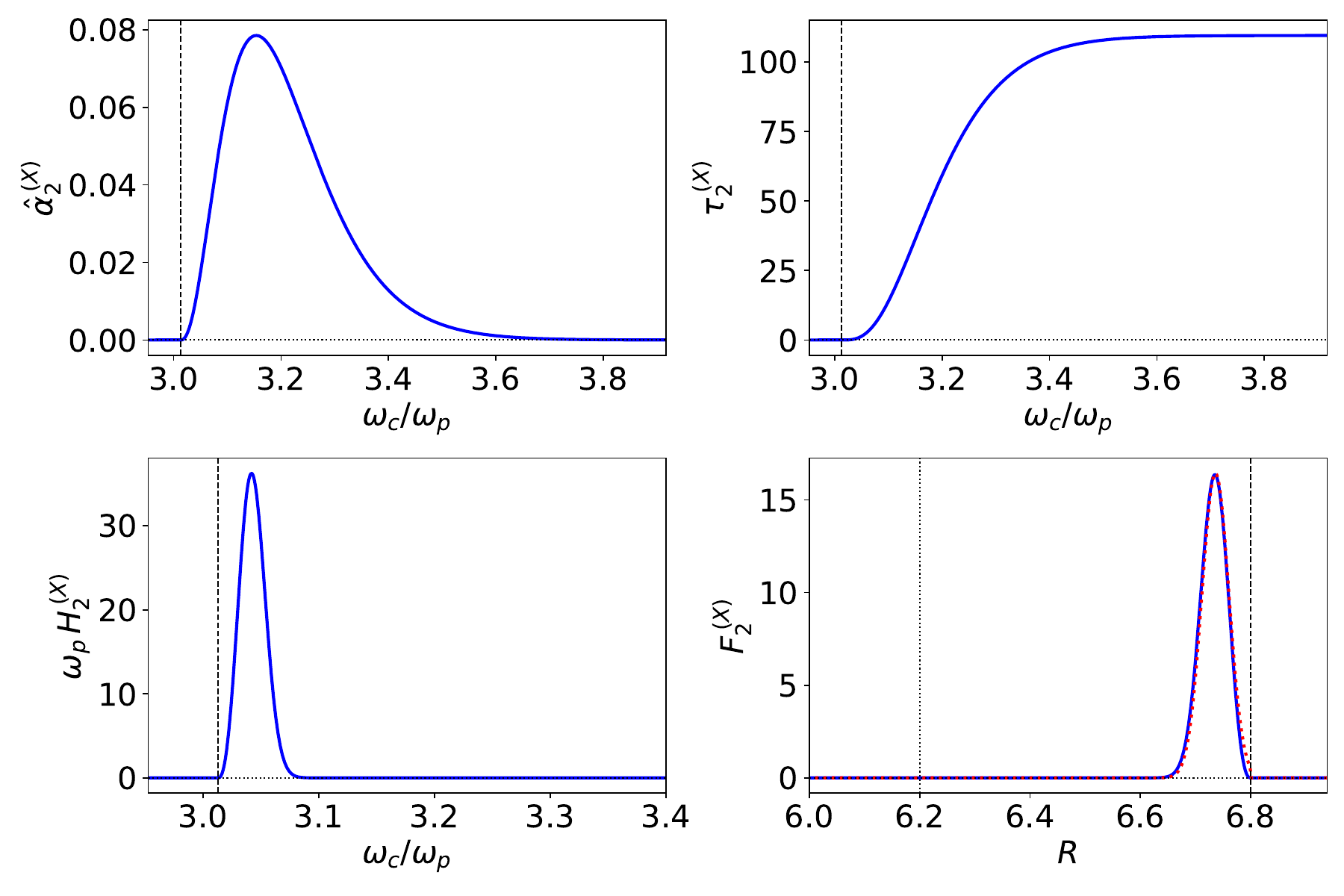}}
\caption{The normalized  absorption coefficient (top left), optical depth (top right), normalized
spectral convolution function (bottom left), and spatial convolution function (bottom right), for 2nd  harmonic X-mode ECE from an 
ITER-like plasma characterized by  $T_e=10\,{\rm keV}$, $n_e=2.5\times  10^{19}\,{\rm m}^{-3}$ (at the 2nd harmonic cyclotron resonance), $B_0=5.3\,{\rm T}$, $R_0=6.2\,{\rm m}$, and $R_\omega= 6.8\,{\rm m}$.
The vertical dashed lines show the location of the 2nd harmonic cyclotron resonance.  The vertical dotted line (in the bottom right panel) shows the location of the  magnetic axis.
The dotted red curve in the bottom right panel is a  fit to the true spatial convolution function (shown in blue).
\label{Xmode}}
\end{figure}

\end{document}